\documentclass[zpreprint,zbstnp]{zeus_paper}

\usepackage[english]{babel}

\chardef\usc=95
\chardef\til=126
\catcode`\@=11 % @ signs are now treated as letters
\DeclareRobustCommand\xdotspace{\futurelet\@let@token\@xdotspace}
\def\@xdotspace{%
  \ifx\@let@token.\else
  \ifx\@let@token\bgroup.\else
  \ifx\@let@token\egroup.\else
  \ifx\@let@token\/.\else
  \ifx\@let@token\ .\else
  \ifx\@let@token~.\else
  \ifx\@let@token!.\else
  \ifx\@let@token,.\else
  \ifx\@let@token:.\else
  \ifx\@let@token;.\else
  \ifx\@let@token?.\else
  \ifx\@let@token/.\else
  \ifx\@let@token'.\else
  \ifx\@let@token).\else
  \ifx\@let@token-.\else
  \ifx\@let@token\@xobeysp.\else
  \ifx\@let@token\space.\else
  \ifx\@let@token\@sptoken.\else
   .\space
   \fi\fi\fi\fi\fi\fi\fi\fi\fi\fi\fi\fi\fi\fi\fi\fi\fi\fi}
\catcode`\@=12 % @ signs are no longer letters

\newcommand{\stru}[2]{%
   \relax\ifmmode\hbox{\vrule height#1 depth#2 width0pt}%
   \else\vrule height#1 depth#2 width0pt\fi}
\newcommand{\Ronum}[1]{\uppercase\expandafter{\romannumeral#1}}
\newcommand{\ronum}[1]{\expandafter{\romannumeral#1}}
\DeclareRobustCommand{\LaTeXZ}{%
  \LaTeX\kern-.05em4\kern-.1em
  {\raisebox{-0.2ex}{$\scriptstyle\text{ZEUS}$}}\xspace}
\DeclareMathAlphabet{\mathbf}{OT1}{cmr}{bx}{sl}
\newcommand{\eVdist}{\kern-0.06667em}

\newcommand{\Gev}{{\text{Ge}\eVdist\text{V\/}}}

\newcommand{\pb}{\,\text{pb}}

\newcommand{\Tesla}{\,\text{T}}

\newcommand{\slashfrac}[2]{%
  \raisebox{0.5ex}{\ensuremath #1}\kern-0.12em/\kern-0.08em
  \raisebox{-.8ex}{\ensuremath #2}}
\newcommand{\sqr}[3]{%
    {\vcenter{\hrule height.#3ex\hbox{\vrule width.#2ex height#1ex
     \kern#1ex\vrule width.#3ex}\hrule height.#2ex}}}

\catcode`\@=11 % @ signs are now treated as letters
\newcommand{\parenbar}{\mathpalette\p@renb@r}
\def\p@renb@r#1#2{\vbox{%
  \ifx#1\scriptscriptstyle \dimen@.7em\dimen@ii.2em\else
  \ifx#1\scriptstyle \dimen@.8em\dimen@ii.25em\else
  \dimen@1em\dimen@ii.4em\fi\fi \offinterlineskip
  \ialign{\hfill##\hfill\cr
    \vbox{\hrule width\dimen@ii}\cr
    \noalign{\vskip-.3ex}%
    \hbox to\dimen@{$\mathchar300\hfil\mathchar301$}\cr
    \noalign{\vskip-.3ex}%
    $#1#2$\cr}}}
\catcode`\@=12 % @ signs are no longer letters

\newcommand{\gh}{{\gamma_h}}

\newcommand{\IP}{{\rm I$\kern-0.01667em$P}\xspace}

\mathchardef\qsm=63
\mathchardef\pls=43
\mathchardef\mns=512
\mathchardef\plm=518
\mathchardef\eql=61
\mathchardef\smallleft=300
\mathchardef\smallright=301
\mathchardef\les=316
\mathchardef\gre=318
\mathchardef\leq=532
\mathchardef\grq=533
\catcode`\@=11 % @ signs are now treated as letters
\newcounter{pict@width}
\newcounter{pict@height}
\newlength{\pict@scale}
\newcommand{\psfigadd}[4]{%
\setcounter{pict@width}{1*\ratio{#2+\pict@scale/2}{\pict@scale}}
\setcounter{pict@height}{1*\ratio{#3+\pict@scale/2}{\pict@scale}}
\setlength{\unitlength}{\pict@scale}
\hbox to #2{\hspace{-\fill}\begin{picture}(\thepict@width,\thepict@height)
\put(0,0){\psfig{figure=#1,width=#2,height=#3,clip=}}
\SetScale{0.283466457}
\SetWidth{1.763889}
{#4}
\end{picture}}
}
\newcounter{pict@widthfst}
\newcounter{pict@widthscd}
\newcounter{pict@widthtot}
\newcommand{\psfigaddtwo}[7]{%
\setcounter{pict@widthfst}{1*\ratio{#2+\pict@scale/2}{\pict@scale}}
\setcounter{pict@widthscd}{1*\ratio{#2+#4+\pict@scale/2}{\pict@scale}}
\setcounter{pict@widthtot}{1*\ratio{#2+#4+#6+\pict@scale/2}{\pict@scale}}
\setcounter{pict@height}{1*\ratio{#3+\pict@scale/2}{\pict@scale}}
\setlength{\unitlength}{\pict@scale}
\hbox{\hspace{-\fill}\begin{picture}(\thepict@widthtot,\thepict@height)
\put(0,0){\psfig{figure=#1,width=#2,height=#3,clip=}}
\put(\thepict@widthscd,0){\psfig{figure=#5,width=#6,height=#3,clip=}}
\SetScale{0.283466457}
\SetWidth{1.763889}
{#7}
\end{picture}}
}
\newcommand{\psfigror}[4]{%
\setcounter{pict@width}{1*\ratio{#2+\pict@scale/2}{\pict@scale}}
\setcounter{pict@height}{1*\ratio{#3+\pict@scale/2}{\pict@scale}}
\setlength{\unitlength}{\pict@scale}
\hbox{\begin{picture}(\thepict@width,\thepict@height)
\put(0,\thepict@height){\psfig{figure=#1,width=#3,height=#2,clip=,angle=270}}
\SetScale{0.283466457}
\SetWidth{1.763889}
{#4}
\end{picture}}
}
\newcommand{\psfigrol}[4]{%
\setcounter{pict@width}{1*\ratio{#2+\pict@scale/2}{\pict@scale}}
\setcounter{pict@height}{1*\ratio{#3+\pict@scale/2}{\pict@scale}}
\setlength{\unitlength}{\pict@scale}
\hbox{\begin{picture}(\thepict@width,\thepict@height)
\put(0,0){\psfig{figure=#1,width=#3,height=#2,clip=,angle=90}}
\SetScale{0.283466457}
\SetWidth{1.763889}
{#4}
\end{picture}}
}
\catcode`\@=12 % @ signs are no longer letters
\newlength\listtextwidth

\catcode`\@=11 % @ signs are now treated as letters
\newlength{\@tabfninsert}
\newlength{\@tabfnwidth}
\newcommand{\tabfootnote}[2]{%
  \setlength{\@tabfninsert}{0.8em}
  \setlength{\@tabfnwidth}{\textwidth}
  \addtolength{\@tabfnwidth}{-\@tabfninsert}
  \addtolength{\@tabfnwidth}{-0.4em}
  \noindent\makebox[\@tabfninsert][r]{\footnotesize$^{#1}$\hfil}\hfill%
  \parbox[t]{\@tabfnwidth}{\footnotesize #2\hfill}}
\catcode`\@=12 % @ signs are no longer letters

\def\etjet{E_T^{\rm jet}}
\def\etajet{\eta^{\rm jet}}

\def\q2{Q^2}

\def\pb1{pb$^{-1}$}
\def\g2{GeV$^2$}                                                                    
\def\gh{\gamma_h} 
\def\escat{E_e^{\prime}}

\def\kt{k_T}
 
\def\as{\alpha_s}
\def\asz{\alpha_s(M_Z)}
\def\oalphas2{{\cal O}(\alpha\as^2)}

\def\oas{{\cal O}(\as)}
\def\oasz{{\cal O}(\as^0)}
\def\oaso{{\cal O}(\as^1)}
\def\oass{{\cal O}(\as^2)}
\def\colab#1{#1 Coll.}

\def\citeDGLAP{{\cite{%
sovjnp:15:438,*sovjnp:20:94,*parisi:1976,*jetp:46:641,*np:b126:298%
}}\xspace}

\def\citeMUELLER{{\cite{%
np:c18:125,*jp:g17:1443%
}}\xspace}

\def\citeFORWARD{{\cite{%
np:c18:125,*jp:g17:1443,zfp:c54:645,*pl:b287:254,*pl:b278:363%
}}\xspace}
\def\citeDJANGO{{\cite{%
cpc:81:381,*spi:www:djangoh11%
}}\xspace}
\def\citeHERACL{{\cite{%
cpc:69:155
}}\xspace}
\def\citeLEPTO{{\cite{%
cpc:101:108
}}\xspace}
\def\citeARI{{\cite{%
cpc:71:15,*zp:c65:285%
}}\xspace}
\def\citeDISENT{{\cite{%
np:b485:291
}}\xspace}
\def\citeKTCLUS{{\cite{%
np:b406:187
}}\xspace}
\def\citeKTELLSOP{{\cite{%
pr:d48:3160
}}\xspace}
\def\citeCDM{{\cite{%
pl:b165:147,*pl:b175:453,*np:b306:746,*zfp:c43:625%
}}\xspace}
\def\citeLUND{{\cite{%
prep:97:31
}}\xspace}
\def\citeMRST99{{\cite{%
epj:c4:463,*epj:c14:133%
}}\xspace}
\def\citeCTEQ6{{\cite{%
hepph0201195
}}\xspace}
\def\citeDA{{\cite{%
proc:hera:1991:23
}}\xspace}

\begin{document}

\prepnum{{DESY--05--017}}

\title{
Forward jet production in deep inelastic  \\
$ep$ scattering and low-$x$ parton dynamics \\
at HERA
}                                                       
                    
\author{ZEUS Collaboration}
\date{January, 2005}

\abstract{Differential inclusive jet cross sections in neutral current deep
inelastic $ep$ scattering have been measured with the ZEUS detector using
an integrated luminosity of 38.7~\pb1.  The jets have been identified
using the $\kt$ cluster algorithm in the longitudinally invariant
inclusive mode in the laboratory frame; they have been selected with jet
transverse energy, $\etjet$, above 6~GeV and jet pseudorapidity,
$\etajet$, between $-1$ and 3.  Measurements of cross sections as
functions of $\etjet$, $\etajet$, Bjorken $x$ and the photon
virtuality, $\q2$, are presented. Three phase-space regions have been
selected in order to study parton dynamics from the most global to the
most restrictive region of forward-going (close to the proton-beam
direction) jets at low $x$, where the effects of BFKL evolution might
be present. The measurements have been compared to the predictions of
leading-logarithm parton-shower Monte Carlo models and fixed-order
perturbative QCD calculations. In the forward region, 
${\cal O}(\alpha_s^1)$ QCD calculations underestimate the data up to
an order of magnitude at low $x$. An improved description of the data
in this region is obtained by including ${\cal O}(\alpha_s^2)$ QCD
corrections, which account for the lowest-order $\hat{t}$-channel
gluon-exchange diagrams, highlighting the importance of such terms in
the parton dynamics at low $x$.
}

\makezeustitle

\def\3{\ss}

\pagenumbering{Roman}

\begin{center}
{                      \Large  The ZEUS Collaboration              }
\end{center}

  S.~Chekanov,
  M.~Derrick,
  S.~Magill,
  S.~Miglioranzi$^{   1}$,
  B.~Musgrave,
  \mbox{J.~Repond},
  R.~Yoshida\\
  {\it Argonne National Laboratory, Argonne, Illinois 60439-4815},
  USA~$^{n}$
\par \filbreak

  M.C.K.~Mattingly \\
  {\it Andrews University, Berrien Springs, Michigan 49104-0380}, 
  USA
\par \filbreak

  N.~Pavel, A.G.~Yag\"ues Molina \\
  {\it Institut f\"ur Physik der Humboldt-Universit\"at zu Berlin,
    Berlin, Germany}
% ...

\newpage

$^{\    1}$ also affiliated with University College London, UK \\
% ...

\newpage

\begin{tabular}[h]{rp{14cm}}
$^{a}$ &  supported by the Natural Sciences and Engineering Research
Council of Canada (NSERC) \\
$^{b}$ &  supported by the German Federal Ministry for Education and
Research (BMBF), under contract numbers HZ1GUA 2, HZ1GUB 0, HZ1PDA 5,
HZ1VFA 5\\
$^{c}$ &  supported in part by the MINERVA Gesellschaft f\"ur
Forschung GmbH, the Israel Science Foundation (grant no. 293/02-11.2),
the U.S.-Israel Binational Science Foundation and the Benozyio Center
for High Energy Physics\\
$^{d}$ &  supported by the German-Israeli Foundation and the Israel
Science Foundation\\
$^{e}$ &  supported by the Italian National Institute for Nuclear
Physics (INFN) \\
$^{f}$ &  supported by the Japanese Ministry of Education, Culture,
Sports, Science and Technology (MEXT) and its grants for Scientific
Research\\
$^{g}$ &  supported by the Korean Ministry of Education and Korea
Science and Engineering Foundation\\
$^{h}$ &  supported by the Netherlands Foundation for Research on
Matter (FOM)\\
$^{i}$ &  supported by the Polish State Committee for Scientific
Research, grant no. 620/E-77/SPB/DESY/P-03/DZ 117/2003-2005 and grant
no. 1P03B07427/2004-2006\\
$^{j}$ &  partially supported by the German Federal Ministry for
Education and Research (BMBF)\\
$^{k}$ &  supported by RF Presidential grant N 1685.2003.2 for the
leading scientific schools and by the Russian Ministry of Education
and Science through its grant for Scientific Research on High Energy
Physics\\
$^{l}$ &  supported by the Spanish Ministry of Education and Science
through funds provided by CICYT\\
$^{m}$ &  supported by the Particle Physics and Astronomy Research
Council, UK\\
$^{n}$ &  supported by the US Department of Energy\\
$^{o}$ &  supported by the US National Science Foundation\\
$^{p}$ &  supported by the Polish Ministry of Scientific Research and
Information Technology, grant no. 112/E-356/SPUB/DESY/P-03/DZ
116/2003-2005 and 1 P03B 065 27\\
$^{q}$ &  supported by FNRS and its associated funds (IISN and FRIA)
and by an Inter-University Attraction Poles Programme subsidised by
the Belgian Federal Science Policy Office\\
\end{tabular}

\newpage

\pagenumbering{arabic} 
\pagestyle{plain}

\section{Introduction}
\label{sec-int}

Deep inelastic lepton scattering (DIS) off protons provides
information on the parton distribution functions (PDFs) of the
proton. For example, inclusive measurements of the cross section for
the reaction $e + p \rightarrow e + {\rm X}$ as a function of the
virtuality of the exchanged boson, $\q2$, and of the Bjorken-$x$
scaling variable, $x$, have been used to determine $F^p_2(x,Q^2)$
which, in turn, is analysed in a theoretical context to extract the
proton PDFs. Perturbative QCD in the next-to-leading-order (NLO)
approximation has been widely used to perform such extraction and to
test the extent to which it is able to describe the data. Perturbative
QCD can predict only the evolution of the PDFs in $Q^2$; several
approximations have been developed depending on the expected
importance of the different terms in the perturbative expansion. In
the standard approach (DGLAP \citeDGLAP), the evolution equations sum
up all leading double logarithms in $\ln\q2 \cdot \ln{1/x}$ along with
single logarithms in $\ln\q2$ and are expected to be valid for $x$ not
too small. At low $x$, a better approximation is expected to be
provided by the BFKL formalism~\cite{jetp:45:199,*sovjnp:28:822} in
which the evolution equations sum up all leading double logarithms
along with single logarithms in $\ln{1/x}$.

The DGLAP evolution equations have been tested extensively at
HERA~\cite{desy-03-214,pl:b547:164,pr:d67:012007,pl:b507:70,epj:c30:1,pl:b542:193,pl:b515:17},
and were found to describe the data, in general, very well. In particular,
the striking rise of the measured $F^p_2(x,Q^2)$ at HERA with
decreasing $x$ can be accomodated in the DGLAP approach. On the other
hand, the inclusive character of $F_2$ together with the dependence of
the DGLAP predictions on the choice of the input form of the
non-perturbative PDFs at $Q^2=Q^2_0$ may obscure the underlying
dynamics at low $x$. In order to probe the parton dynamics at low $x$,
measurements of the partonic final state that highlight the
differences predicted by the BFKL and DGLAP formalisms were
suggested~\citeMUELLER.

\begin{figure}[p]
\setlength{\unitlength}{1.0cm}
\begin{picture} (19.0,18.50)
\put (1.0,5.0) {\epsfig{file=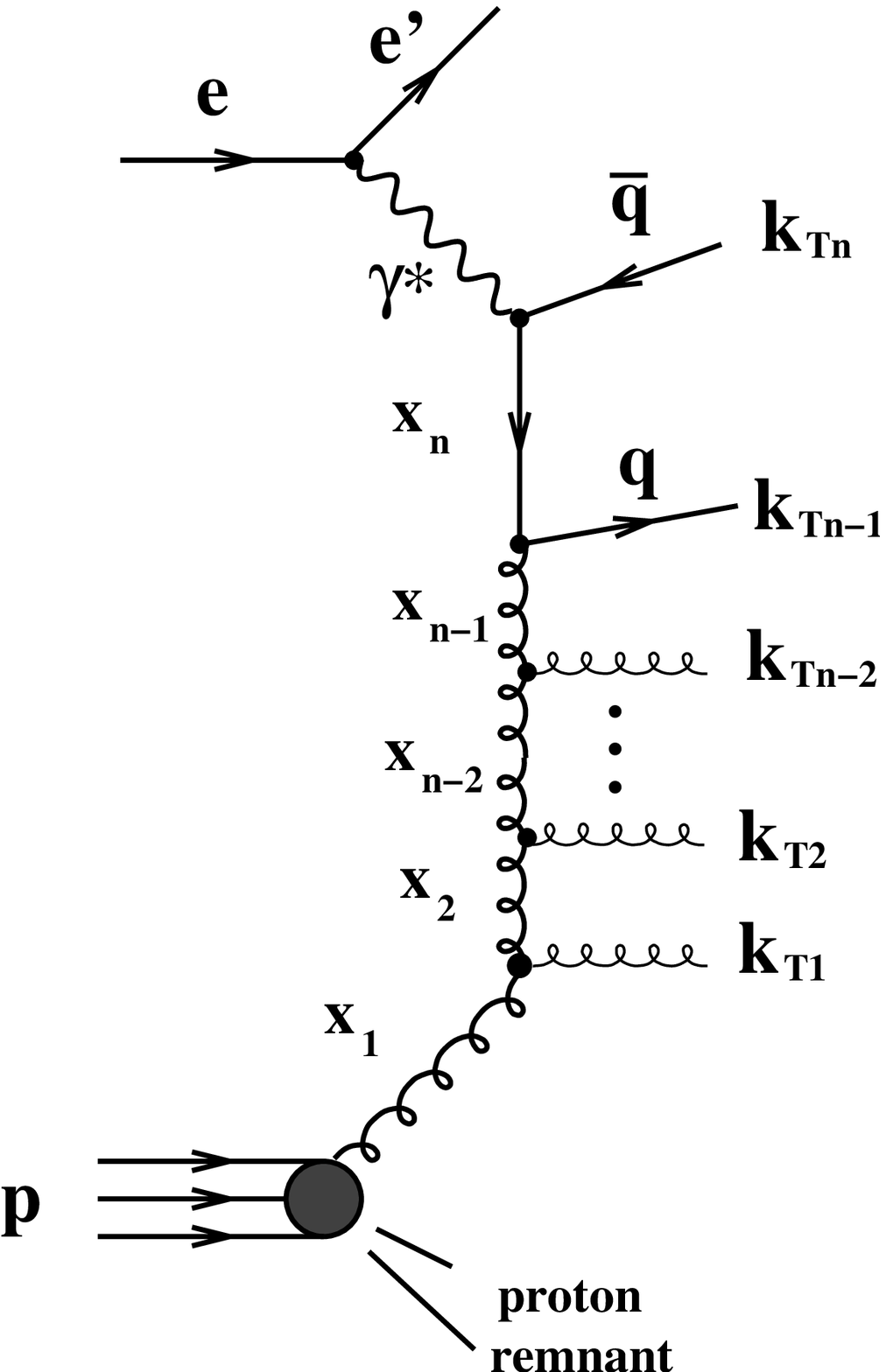,width=6cm}}
\hspace{2cm}
\put (10.0,5.0) {\epsfig{file=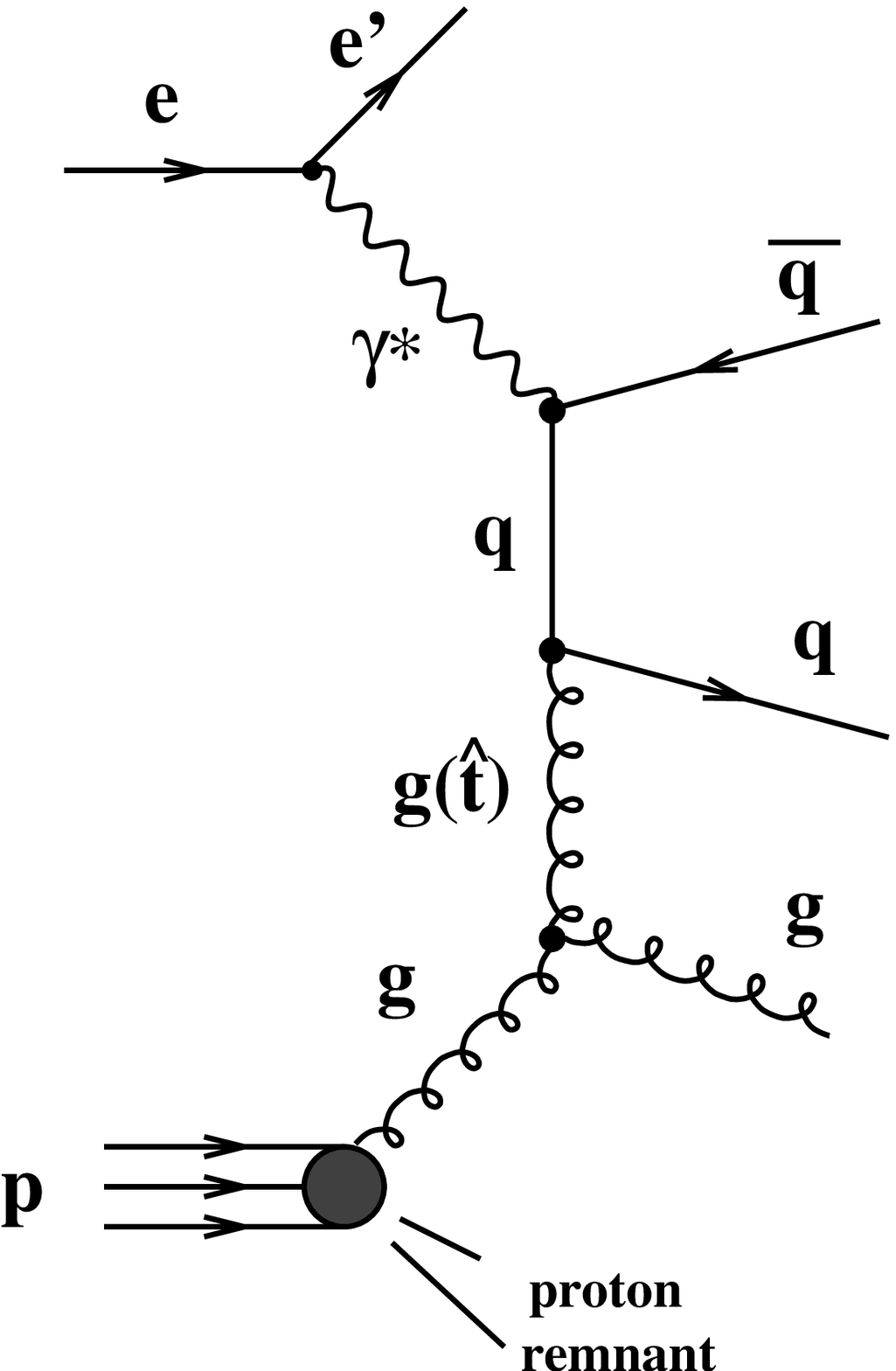,width=6cm}}
\put (4.0,4.0){\bf\small (a)}
\put (13.0,4.0){\bf\small (b)}
\end{picture}
\vspace{-3.0cm}
\caption{
(a) Gluon-ladder Feynman diagram.  In DGLAP evolution, the final-state
partons are ordered in transverse energy, 
$k_{T,n}^2 > k_{T,n-1}^2 > k_{T,1}^2$. In BFKL, the partons are
emitted without any ordering in $k_T$ along the ladder. (b) Example of
Feynman diagram with $\hat{t}$-channel gluon exchange at 
${\cal O}(\alpha_s^2)$.
}
\label{fig-gluon}
\end{figure}

In the DGLAP formalism, the parton cascade that results from the hard
scattering of the virtual photon with a parton from the proton is
ordered in parton virtuality.  This ordering along the parton ladder
implies an ordering in transverse energy of the partons, $E_T$, with
the parton participating in the hard scatter having the highest
transverse energy. In the BFKL formalism, there is no strict ordering
in virtuality or transverse energy (see Fig.~\ref{fig-gluon}a). Since
the partons emitted at the bottom of the ladder are closest in
rapidity to the outgoing proton, they manifest themselves as
forward\footnote{The ZEUS coordinate system is a right-handed
  Cartesian system, with the $Z$ axis pointing in the proton beam
  direction, referred to as the "forward direction", and the $X$ axis
  pointing towards the centre of HERA.  The coordinate origin is at
  the nominal interaction point.} jets. BFKL evolution predicts that a
larger fraction of small $x$ events will contain high-$E_T$ forward
jets than is predicted by DGLAP \citeFORWARD.
 
In previous studies of forward jets in
DIS~\cite{epj:c6:239,pl:b474:223,np:b538:3}, the data were compared to
Monte Carlo simulations which model higher-order parton emissions
using the DGLAP approach. These models do not describe the
data. However, it was possible to obtain a better description of the
data by adding a second $E_T$-ordered parton cascade on the photon
side which is evolved according to the DGLAP equations; this
resolved-photon contribution \cite{cpc:86:147} leads to parton-parton
scattering which can give rise to the production of high-$E_T$ jets
anywhere along the (double) ladder between the photon and the
proton. The calculations based on the Colour Dipole Model (CDM)
\citeCDM , which include parton emissions not ordered in transverse
energy, also described the data. In a more recent analysis
\cite{pl:b542:193}, fixed-order QCD calculations were compared to the
data.  The predictions fail to describe the measurements in the most
forward region at low $E_T$ and $\q2$.

In this paper, measurements of differential inclusive jet cross sections
in deep inelastic scattering are presented in three different phase-space
regions, from the most global to the most restrictive region, where the
contribution of events exhibiting BFKL characteristics should be enhanced.
A novel method is introduced (see Section~\ref{sec-ps}) to increase the
sensitivity to additional parton radiation in the forward region while
extending the region in $x$ towards lower values. The jets were
reconstructed using the $\kt$ cluster algorithm~\citeKTCLUS in the
longitudinally invariant mode~\citeKTELLSOP, instead of the cone
algorithm used in previous studies
\cite{epj:c6:239,pl:b474:223,np:b538:3}, which allows a reduction of
the theoretical uncertainty associated with matching the experimental
and theoretical jet algorithms.  Inclusive jet cross sections were
measured as functions of the jet transverse energy, $\etjet$,
pseudorapidity, $\etajet$, and the event variables $\q2$ and $x$. The
effects of higher-order terms in the parton cascade were explored by
comparing the data to fixed-order QCD predictions using current
parametrisations of the proton PDFs based on DGLAP evolution. In
addition, the predictions of a leading-logarithm parton-shower model
based on DGLAP evolution and those of an implementation of the
colour-dipole model were also compared to the data.

\section{Theoretical expectations and phase-space definitions}
\label{sec-ps}

For a given $e^+p$ centre-of-mass energy, $\sqrt{s}$, the cross section
for neutral current (NC) deep inelastic $ep$ scattering,
$e^+ p \rightarrow e^+ + {\rm X}$, depends on two independent kinematic
variables, which are chosen to be $Q^2$ and the Bjorken scaling variable,
$x$, where $Q^2=-q^2$ and $x = Q^2/(2 P\cdot q)$; $P$ ($q$)
is the four-momentum of the incoming proton (exchanged virtual boson,
$V^*$, with $V=\gamma$ or $Z^0$). Other variables used to define the
kinematics of the events are $y=Q^2/(x s)$ and $\gh$, defined by
$\cos{\gh}=((1-y)x E_p - y E_e)/((1-y)x E_p + y E_e)$, where
$E_p$ ($E_e$) is the energy of the incoming proton (positron).

Jet production in NC DIS at $\oasz$ proceeds via $V^* q \rightarrow q$;
this process is referred to as being of quark-parton-model (QPM) type.
The hadronic final state consists of a single
jet emerging at polar angle $\gh$ and balancing the transverse momentum
of the scattered $e^+$ plus the remnant of the proton. The NLO QCD
corrections of $\oaso$ consist of one-loop corrections to the process
$V^* q \rightarrow q$ and the tree-level processes of boson-gluon
fusion (BGF, $V^* g \rightarrow q\bar{q}$) and QCD-Compton (QCDC, 
$V^*q \rightarrow q g$). In BGF and QCDC, when the two final-state
partons are sufficiently separated from each other, the hadronic final
state consists of two jets plus the remnant of the proton.

The predictions of fixed-order QCD calculations convoluted with PDFs
extracted using the DGLAP equations have the following features for
inclusive jet production: a dominant contribution ($\oasz$) from
single-jet events with $\theta^{\rm jet}=\gh$ and a $\oaso$-suppressed
contribution from dijet events. Since at low values of $x$ the variable
$\gh$ points toward the rear direction, the production of forward (having
positive values of $\etajet$) jets is suppressed. In this region BFKL
predicts a higher forward-jet cross section than DGLAP. This effect can be
further enhanced by suppressing the evolution in $Q^2$ by
requiring $(\etjet)^2 \sim Q^2$.

Experimental studies of QCD using jet production in NC DIS at HERA are
often performed in the Breit frame \cite{feynman:1972:photon}. The
analysis presented here was instead performed in the laboratory frame
for two reasons. First, such an analysis provides access to low values
of $x$: the requirement of a jet in the Breit frame with a given
$\etjet$ would demand a larger fraction of the proton's momentum than
that of a jet (with the same $\etjet$) in the laboratory frame. It is
noted that, in the Breit frame, the exchanged virtual boson collides
head-on with the proton and, therefore, the transverse momentum of a
jet must be balanced by other jet(s). Second, the application of the
jet algorithm in the laboratory frame benefits from the increased
resolution for identifying jets in the forward region of the
detector. Jet cross sections in the laboratory frame are theoretically
well defined and NLO QCD calculations for such observables are well
behaved \cite{jp:g25:1473}.

To investigate the NLO QCD predictions in detail, three phase-space
regions of inclusive jet production have been studied. The first
region, called ``global'', was designed to be as inclusive as possible
to keep the theoretical uncertainties small. This region was defined
by the conditions:
\begin{itemize}
\item $Q^2 > 25~\rm{GeV^2}$;
\item $y > 0.04$;
\item $\escat > 10~\rm{GeV}$, where $\escat$ is the energy of the
  scattered positron;
\item at least one jet with $\etjet > 6~\rm{GeV}$ and $-1 < \etajet < 3$.
\end{itemize}
This phase-space region is expected to be dominated by QPM-type events.

A second phase-space region, called ``BFKL'', was defined by the following
additional conditions:
\begin{itemize}
\item $\cos\gh < 0$;
\item at least one jet with $0 < \etajet < 3$ and 
  $0.5 < \frac{(\etjet)^2}{Q^2} < 2$.
\end{itemize}
The combination of the requirements $\gh > 90^{\circ}$ and 
$\theta^{\rm jet}< 90^{\circ}$ suppresses the contribution from
QPM-type events. This phase-space region is expected to be dominated
by multi-jet events. The enhancement of the contribution from
multi-jet events is done without an explicit requirement on the number
of jets and, thereby, keeping events at low values of $x$. The
requirement on $(\etjet)^2/Q^2$ restricts the jet kinematics to the
region where the BFKL effects are expected to be large.

A third phase-space region, called ``forward BFKL'', was designed to
investigate events with a very forward-going jet and was defined by
requiring, in addition to the aforementioned cuts, at least one jet
with $2 < \etajet < 3$.

\section{Experimental set-up}
\label{sec-exp}

The data sample used in this analysis was collected with the ZEUS
detector at HERA and corresponds to an integrated luminosity of 
$38.7 \pm 0.6~${\rm \pb1}. During 1996-1997, HERA operated with
protons of energy $E_p=820$~GeV and positrons of energy
$E_e=27.5$~GeV. A detailed description of the ZEUS detector can be
found elsewhere~\cite{pl:b293:465,zeus:1993:bluebook}. A brief outline
of the components that are most relevant for this analysis is given
here.

Charged particle tracks are reconstructed in the central tracking
detector (CTD)~\cite{nim:a279:290,*npps:b32:181,*nim:a338:254}, which
operates in a magnetic field of $1.43\Tesla$ provided by a thin
superconducting solenoid. The CTD consists of 72~cylindrical
drift-chamber layers, organised in nine superlayers covering the
polar-angle region \mbox{$15^\circ<\theta<164^\circ$}. The
transverse-momentum resolution for full-length tracks can be
parameterised as 
$\sigma(p_T)/p_T=0.0058p_T\oplus0.0065\oplus0.0014/p_T$, with $p_T$ in
$\Gev$. The tracking system was used to measure the interaction vertex
with a typical resolution along (transverse to) the beam direction of
0.4~(0.1)~cm and to cross-check the energy scale of the calorimeter.
 
The high-resolution uranium--scintillator calorimeter
(CAL)~\cite{nim:a309:77,*nim:a309:101,*nim:a321:356,*nim:a336:23} covers
$99.7\%$ of the total solid angle and consists
of three parts: the forward (FCAL, $2.6^{\circ}<\theta<36.7^{\circ}$), the
barrel (BCAL, $36.7^{\circ}<\theta<129.1^{\circ}$) and the rear
(RCAL, $129.1^{\circ}<\theta<176.2^{\circ}$) calorimeters. Each part is
subdivided transversely into towers and longitudinally into one
electromagnetic section (EMC) and either one (in RCAL) or two (in BCAL
and FCAL) hadronic sections (HAC). The smallest subdivision of the
calorimeter is called a cell. Under test-beam conditions, the CAL
single-particle relative energy resolutions were 
$\sigma(E)/E=0.18/\sqrt E$ for electrons and
$\sigma(E)/E=0.35/\sqrt E$ for hadrons, with $E$ in GeV.
 
The luminosity was measured from the rate of the bremsstrahlung
process $ep\rightarrow e\gamma p$. The resulting small-angle
energetic photons were measured by the luminosity
monitor~\cite{desy-92-066,*zfp:c63:391,*acpp:b32:2025}, a
lead-scintillator calorimeter placed in the HERA tunnel at $Z=-107$~m.

\section{Data selection and jet identification}
\label{sec-data}

A three-level trigger was used to select events online
\cite{zeus:1993:bluebook}. The NC DIS events were selected offline
using criteria similar to those reported previously
\cite{hep-ex-0208037}.  The main steps are outlined below.

The scattered positron candidate was identified from the pattern of
energy deposits in the CAL. The $\escat$ and polar angle ($\theta_e$)
of the positron candidate were also determined from the CAL
measurements, after correction for energy loss in inactive material in
front of the CAL. The following requirements were imposed on the data
sample:
\begin{itemize}
\item the reconstructed $\q2 > 25$~GeV$^2$;
\item a positron candidate of energy $\escat > 10~{\rm GeV}$.  This cut
  ensured a high and well understood positron-finding efficiency and
  suppressed background from photoproduction events, in which the
  scattered positron escapes undetected in the rear beampipe;
\item the vertex position, determined from CTD tracks, in the range
  $\left|Z_{\rm vtx}\right| < 50~\rm{cm}$ along the beam axis.
  This cut removed background events from non-$ep$ interactions; 
\item $38 < (E-P_Z) < 65~\rm{GeV}$, where $E$ is the total energy measured
  in the CAL, $E=\sum_i E_i$, and $P_Z$ is the $Z$ component of the vector
  ${\bf p} = \sum_i E_i {\bf r_i}$; in both cases the sum runs over
  all CAL cells, $E_i$ is the energy of the CAL cell $i$ and 
  ${\bf r_i}$ is a unit vector along the line joining the
  reconstructed vertex to the geometric centre of the cell $i$. This
  cut removed events with large initial-state radiation and further
  reduced the background from photoproduction events;
\item $y_{e} < 0.95$, where $y_{e} = 1 - \escat(1-\cos{\theta_e})/(2E_e)$. 
  This condition removed events in which fake positron candidates from 
  photoproduction background were found in the FCAL;
\item $y_{\rm JB} > 0.04$, where 
  $y_{\rm JB} = \sum_i E_i (1-\cos\theta_i)/(2E_e)$
  calculated according to the Jacquet-Blondel method
  \cite{proc:epfacility:1979:391}, where the sum runs over all CAL
  cells except those assigned to the scattered positron. The 
  $y_{\rm JB}$ variable is an estimator of $y$ which gives a good
  resolution at low $y$;
\item $p_T^{\rm CAL} / \sqrt{E_T^{\rm CAL}} < 3~\sqrt{\rm{GeV}}$, where
  $p_T^{\rm CAL}$ is the total transverse momentum as measured with
  the CAL and $E_T^{\rm CAL}$ is the total transverse energy in the
  CAL.  This cut removed cosmic rays and beam-related background;
\item $|X| > 14$ cm or $|Y| > 14$ cm, where
  $X$ and $Y$ are the impact positions of the
  positron on the CAL, to avoid the low-acceptance region adjacent to
  the rear beampipe;
\item the energy not associated with the positron candidate within a 
  cone of radius 0.7 units in the pseudorapidity-azimuth ($\eta-\phi$)
  plane around the positron direction was required to be less than
  10\% of the positron energy. This condition removed photoproduction
  and DIS events in which a part of a jet was incorrectly identified
  as the scattered positron.
\end{itemize}

The kinematic variables $\q2$ and $x$ were reconstructed using a
combination of the electron and double-angle (DA) methods \citeDA,
depending on which method gave a better resolution of the observed
scattered-positron energy. The angle $\gh$ was reconstructed with the
CAL using:
\begin{equation}
\cos\gh = \frac{{(\sum_i p_{X,i})^2 + (\sum_i p_{Y,i})^2 - (\sum_i (E-p_Z)_i)^2}}
             {{(\sum_i p_{X,i})^2 + (\sum_i p_{Y,i})^2 + (\sum_i (E-p_Z)_i)^2}},
  \nonumber
\end{equation}
where the sum runs over all CAL cells, excluding those associated with the
scattered positron.  
 
The $k_T$ cluster algorithm was used in the longitudinally invariant
inclusive mode to reconstruct jets in the hadronic final state from
the energy deposits in the CAL cells.  The jet algorithm was applied
after excluding those cells associated with the scattered-positron
candidate. The jet search was performed in the $\eta-\phi$ space in
the laboratory frame. The jet variables were defined according to the
Snowmass convention~\cite{proc:snowmass:1990:134}. Jet transverse
energies were corrected for all energy-loss effects, principally in
inactive material, typically about one radiation length, in front of
the CAL. After these corrections to the jet transverse energy, events
with at least one jet satisfying $\etjet > 6$~GeV and $-1 < \etajet <
3$ were included in the global data sample.  The BFKL and forward-BFKL
subsamples were selected using the additional requirements listed in
Section~\ref{sec-ps}.

\section{Monte Carlo simulation}
\label{sec-mc}

Samples of events were generated to determine the response of the
detector to jets of hadrons and to determine the correction factors
necessary to obtain the hadron-level jet cross sections.  The
generated events were passed through the {\sc
  Geant}~3.13-based~\cite{tech:cern-dd-ee-84-1} ZEUS detector- and
trigger-simulation programs \cite{zeus:1993:bluebook}. They were
reconstructed and analysed by the same program chain as the data.

Neutral current DIS events were generated using the {\sc Lepto}~6.5.1
program \citeLEPTO interfaced to {\sc Heracles}~4.6.1 \citeHERACL via
{\sc Djangoh}~1.1 \citeDJANGO.  The {\sc Heracles} program includes
photon and $Z$ exchanges and first-order electroweak radiative
corrections. The QCD cascade was modelled with the CDM as implemented
in the {\sc Ariadne}~4.08 program \citeARI; {\sc Ariadne} simulates
the BGF process in addition.  The CDM treats gluons emitted from
quark-antiquark (diquark) pairs as radiation from a colour dipole
between two partons. This results in partons that are not ordered in
their transverse momenta. Samples of events were also generated using
the model of {\sc Lepto} based on first-order QCD matrix elements plus
parton showers (MEPS). For the generation of the samples with MEPS,
the option for soft-colour interactions was switched
off~\cite{epj:c11:251}. In both cases, fragmentation into hadrons was
performed using the Lund string model \citeLUND as implemented in {\sc
  Jetset}~7.4 \cite{cpc:82:74}.  The CTEQ4M \cite{pr:d55:1280} proton
PDFs were used for all simulations.

The jet identification was performed using the simulated energy
measured in the CAL cells in the same way as for the data.  The same
jet algorithm was also applied to the final-state particles and to the
partons available after the parton shower; the jets found in this way
are referred to as hadronic and partonic jets, respectively.

Electroweak-radiative and hadronisation effects are not at present
included in the fixed-order QCD programs described in
Section~\ref{sec-nlo}. Therefore, samples of Monte Carlo (MC) events
were generated with and without electroweak-radiative effects so that
the data could be corrected. The samples without electroweak-radiative
effects were used to correct the QCD calculations for hadronisation
effects. Since the measurements refer to jets of hadrons, whereas the
QCD calculations refer to partons, the predictions were corrected to
the hadron level using these MC samples. A multiplicative correction
factor, $C_{\rm had}$, was defined as the ratio of cross sections for
jets of hadrons over that for jets of partons, and was computed with
the MC programs. The factor applied to the predictions was the average
of the correction factors obtained with {\sc Ariadne} and {\sc Lepto}.
The uncertainty on the hadronisation correction was taken to be the
absolute difference in the correction factors obtained with {\sc
  Ariadne} and {\sc Lepto}.

\section{Acceptance corrections and systematic uncertainties}
\label{sec:acc}

The {\sc Ariadne} MC samples of events were used to compute the acceptance
corrections.  These correction factors took into account the efficiency of
the trigger, the selection criteria, and the purity and efficiency of the
jet identification, and were generally between 0.8 and 1.2.  The inclusive
cross sections for jets of hadrons were determined by applying bin-by-bin
corrections to the measured distributions.  For this approach to be valid,
the distributions in the data must be well described by the MC simulations
at the detector level.  This condition was in general satisfied by both
{\sc Ariadne} and {\sc Lepto}. The {\sc Lepto} MC samples were used to
compute the systematic uncertainties coming from the parton-shower
simulation.

A study of the main sources contributing to the systematic uncertainties 
of the measurements was performed~\cite{thesis:lammers:2004}. These sources
were:
\begin{itemize}
\item  the parton-shower simulation.  The effect of the treatment of the
   parton shower was estimated using {\sc Lepto} to evaluate the
   acceptance-correction factors.  The difference in the corrected cross
   sections between using {\sc Ariadne} and {\sc Lepto} was taken
   to be the value of systematic uncertainty;
\item  the choice of reconstruction method for the kinematic variables
  $\q2$ and $x$. The difference in the corrected cross sections
  between using the electron and double-angle methods was taken to be
  the value of systematic uncertainty;
\item  the biases introduced by the selection cuts.  The uncertainty due to
   the selection requirements was computed by varying the values of the
   cuts in data and MC. The largest effects were due to the cuts on jet
   transverse energy and hadronic angle.
\end{itemize}

These systematic uncertainties were added in quadrature. The absolute
energy scale of the jets in data events was varied by its uncertainty
of $3\%$~\cite{epj:c23:13}. This uncertainty is highly correlated
between measurements in different bins and is therefore treated
separately. The largest contribution to the overall systematic
uncertainty was due to the uncertainty in the jet energy scale, which
averaged about 5$\%$, but could reach values as high as $20\%$. The
second-largest contribution was due to the choice of parton-shower
simulation, which had effects on the corrected cross section generally
below $5\%$; in the most restrictive phase space, however, the
$\etjet$, $x$ and $\q2$ bins with the fewest events had large
systematic differences. The uncertainty in the luminosity
determination of $1.6\%$ was not included.

\section{QCD calculations}
\label{sec-nlo}

The measurements were compared with QCD predictions evaluated using 
the program {\sc Disent}~\citeDISENT. The calculations were performed in the
$\overline{\rm MS}$ renormalisation and factorisation schemes using a
generalised version of the subtraction method~\citeDISENT. The number
of flavors was set to five; the renormalisation ($\mu_R$) and
factorisation ($\mu_F$) scales were both set to $\mu_R = \mu_F = Q$;
$\alpha_{s}$ was calculated at two loops using 
$\Lambda_{\overline{\rm MS}}^{(5)} = 226$~MeV, which corresponds to
$\asz = 0.1180$.  The CTEQ6~\citeCTEQ6 parameterisations of the proton
PDFs were used. The results obtained with {\sc Disent} were
cross-checked using the program {\sc Disaster}++
\cite{Graudenz:1997sp,*Graudenz:1997gv}; the differences were found to
be less than 1\% in most cases, and never exceeded 3\%.

DISENT allows calculations that sum up to two orders of the
perturbation series. In the global phase-space region, the sample is
dominated by single-jet events. Therefore, the {\sc Disent}
predictions in this region were calculated using the diagrams with
${\cal O} (\alpha_{s}^0)$ and ${\cal O} (\alpha_{s}^1)$. In the BFKL
and forward-BFKL phase-space regions, the samples consist of multijet
events, and so the {\sc Disent} calculations were computed using terms
with ${\cal O}(\alpha_{s}^1)$ and 
${\cal O}(\alpha_{s}^2)$. Perturbative QCD calculations at 
${\cal O}(\alpha_{s}^2)$ can give rise to forward jet production at
low $x$ through diagrams such as that shown in Fig.~\ref{fig-gluon}b,
but no $E_T$-ordering is explicitely imposed among the three
final-state partons.

The following sources of theoretical uncertainties were considered:
\begin{itemize}
\item the choice of renormalisation scale. The uncertainty in the
  calculations arising from the absence of higher-order terms was
  estimated by varying $\mu_R$ by a factor of two up and down. The
  effect on the calculations is between 5 and 50$\%$, depending on the
  phase-space region;
\item the uncertainties in the proton PDFs. The effect of these
  uncertainties in the calculations was estimated by repeating the
  calculations using 40 additional sets from CTEQ6. The resulting
  uncertainty was always below 5$\%$;
\item the choice of factorisation scale. The uncertainty in the
  calculations was estimated by varying $\mu_F$ by a factor of two up
  and down.  The effect on the calculations was usually less than
  5$\%$, except in the global phase-space region where it contributed
  a 20$\%$ uncertainty at low $x$.
\end{itemize}

\section{Results}
The measurements of cross sections differential in $X$, where $X$ is
$\etajet$, $\etjet$, $Q^2$ or $x$, are presented in the three
phase-space regions. The measured cross sections were corrected to
hadron level by the formula:
\begin{equation}
\left(\frac{d\sigma}{dX}\right)_{\rm had} = \frac{N_{\rm Data}}{{\cal L}\cdot \Delta X}\cdot
  \frac{N_{\rm MC}^{\rm had}}{N_{\rm MC}^{\rm det}}\cdot
  \frac{N_{\rm MC}^{\rm no QED}}{N_{\rm MC}^{\rm QED}}
  \nonumber
\end{equation}
where $\Delta X$ is the bin size, $N_{\rm Data}$ are the numbers of
data events, $\cal L$ is the integrated luminosity, 
$N_{\rm MC}^{\rm had}$ ($N_{\rm MC}^{\rm det}$) is the hadron-
(detector-) level MC distribution, $N_{\rm MC}^{\rm QED}$ 
($N_{\rm MC}^{\rm no QED}$) is the hadron-level MC distribution
generated with (without) QED radiation.

The cross sections as functions of $\etajet$ and $\etjet$ are
measurements of every jet in the event, whereas the cross sections as
functions of $Q^2$ and $x$ are event cross sections for the events in
the inclusive jet samples.

\subsection{The global phase-space region}
\begin{figure}[p]
\vfill
\setlength{\unitlength}{1.0cm}
\begin{picture} (18.0,18.0)
\put (-1.0,9.0){\epsfig{file=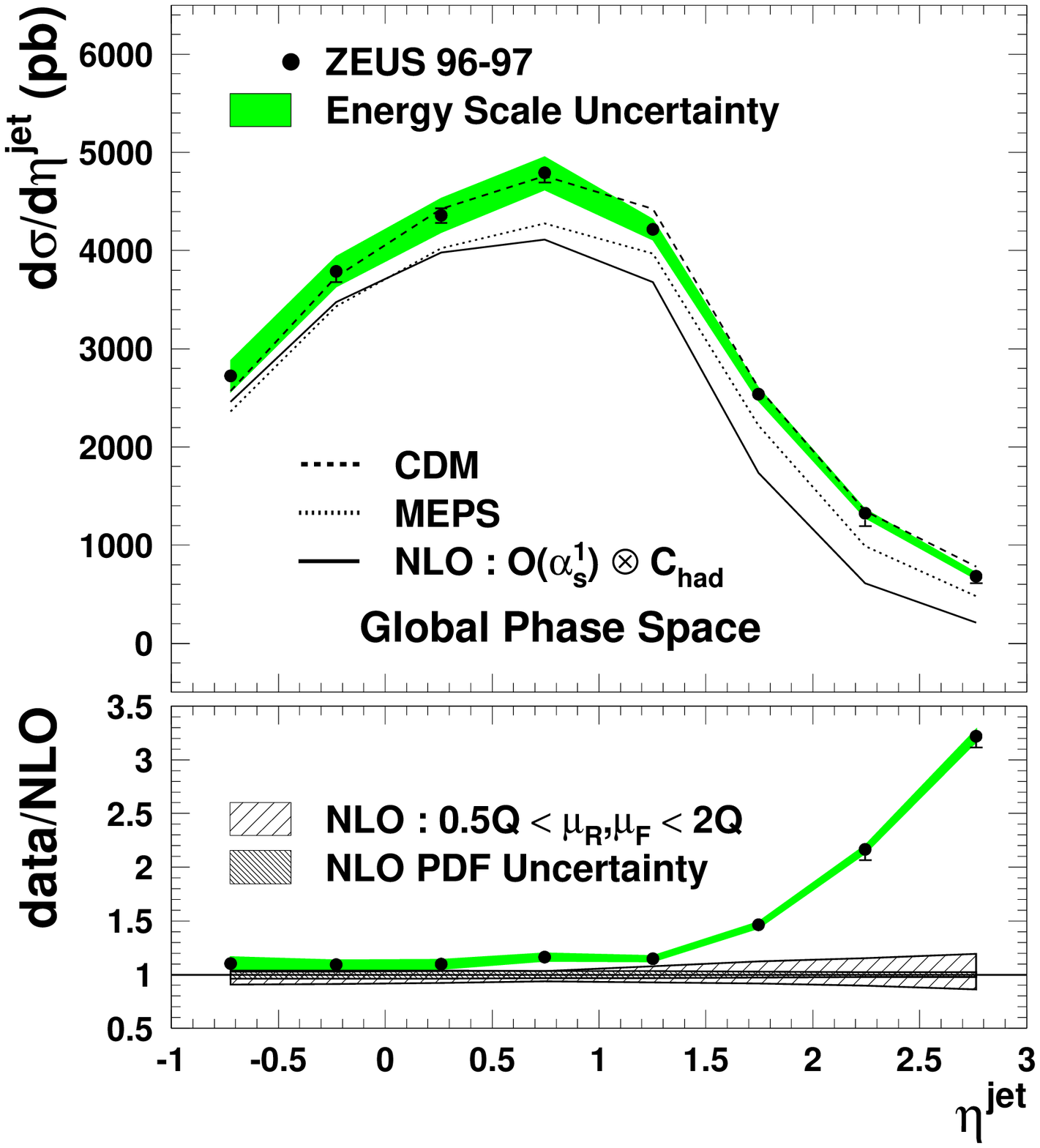,width=10cm}}
\put (8.0,9.0){\epsfig{file=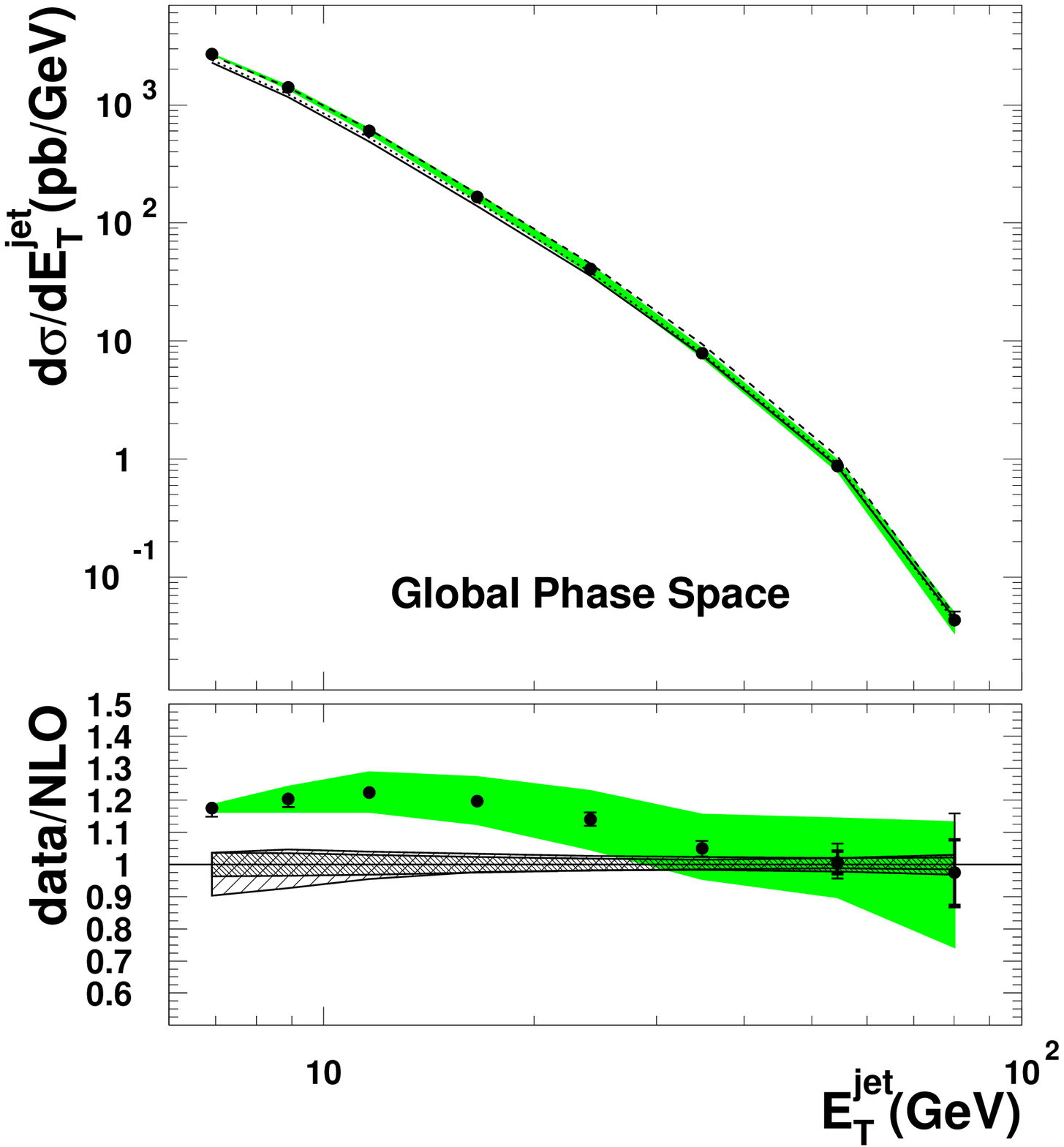,width=10cm}}
\put (-1.0,0.0){\epsfig{file=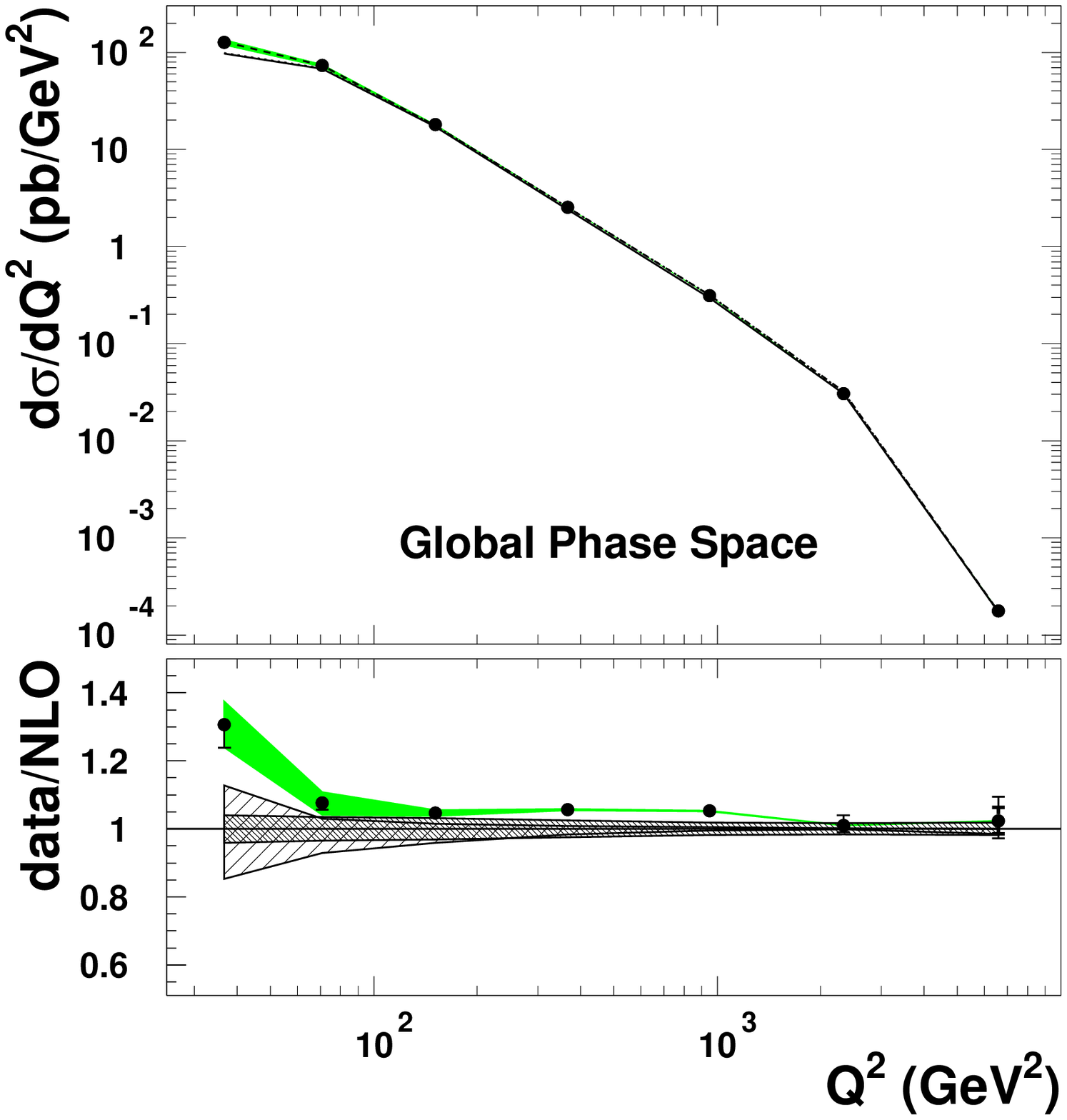,width=10cm}}
\put (8.0,0.0){\epsfig{file=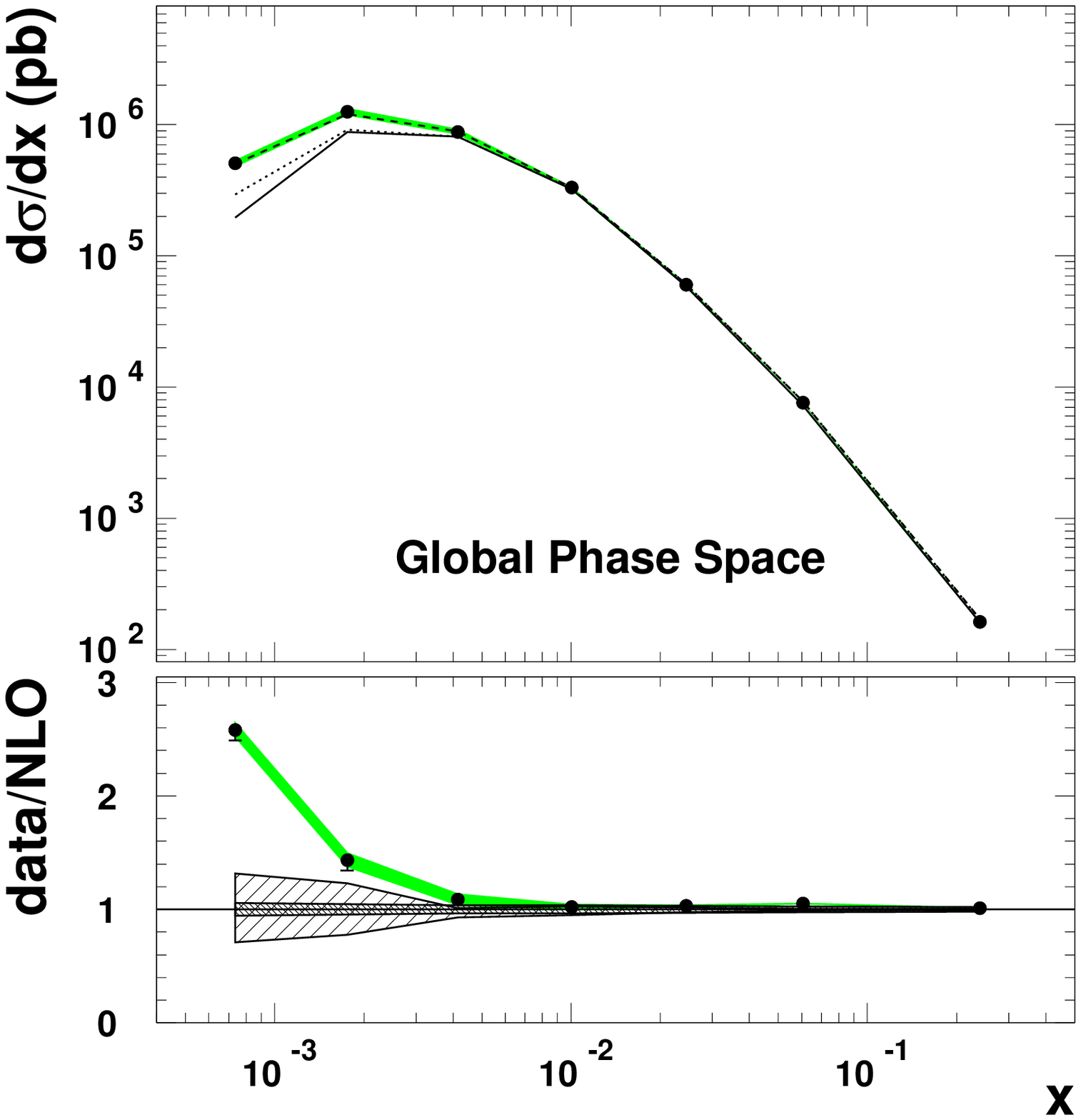,width=10cm}}
\put (6.0,17.0){\bf\small (a)}
\put (15.0,17.0){\bf\small (b)}
\put (6.0,7.5){\bf\small (c)}
\put (15.0,7.5){\bf\small (d)}
\put (7.25,19.0){\bf\huge ZEUS}
\end{picture}
\vspace{-1.0cm}
\caption{
Differential cross sections (dots) in the global phase space for
inclusive jet production in NC DIS with $\etjet > 6$~${\rm GeV}$,
$-1 < \etajet < 3$, $\q2 > 25$~${\rm GeV}^2$ and $y > 0.04$ as
functions of (a) $\etajet$, (b) $\etjet$, (c) $\q2$ and (d) $x$.
The uncertainties are generally smaller than the markers; where visible
the thick error bars represent the statistical uncertainty and the thin
error bars show the statistical and systematic uncertainties added in
quadrature. The uncertainty in the absolute energy scale of the jets is
shown separately as a shaded band. The calculations of CDM
(dashed lines), MEPS (dotted lines) and $\oas$ QCD calculations
(solid lines) are shown. The lower part of each plot shows the ratio
of data to the QCD calculations and the theoretical uncertainties.
}
\label{fig-incl}
\vfill
\end{figure}

The measurements in the global phase-space region are presented in
Fig.~\ref{fig-incl} and Tables 1 to 4. 
The cross section as a function of $\etajet$ is suppressed
in the forward region (high $\etajet$) due to the lower cut on $y$.
The measurements span $x$ values between $0.00074$ and $0.24$.

The MC predictions and fixed-order QCD calculations are compared to the
data in Fig.~\ref{fig-incl}. The prediction of {\sc Ariadne} describes
all data distributions well, whereas the predictions from {\sc Lepto}
are slightly worse, especially at the lowest $x$ values. The
fixed-order QCD calculations describe the data at high $\etjet$, $\q2$
and $x$ values. However, at low values of these variables, the
calculations underestimate the data, and the $\etajet$ distribution is
not described, particularly at high values of $\etajet$.  This excess
of the data with respect to the calculations can be due to
the absence of higher orders, since these calculations are only $\oas$.
The small uncertainty coming from the variation of $\mu_R$ is not expected
to be a reliable estimate of the contributions from higher orders:
contributions from gluon exchange in the $\hat{t}$ channel (as shown
in Fig.~\ref{fig-gluon}b), which become dominant at low $x$
\cite{DelDuca:hepph9707348}, appear only at higher orders, but their
effects cannot be inferred from scale variations of the (lower) terms
considered in the calculation.

\subsection{The BFKL phase-space region}
\begin{figure}[p]
\vfill
\setlength{\unitlength}{1.0cm}
\begin{picture} (18.0,18.0)
\put (-1.0,9.0){\epsfig{file=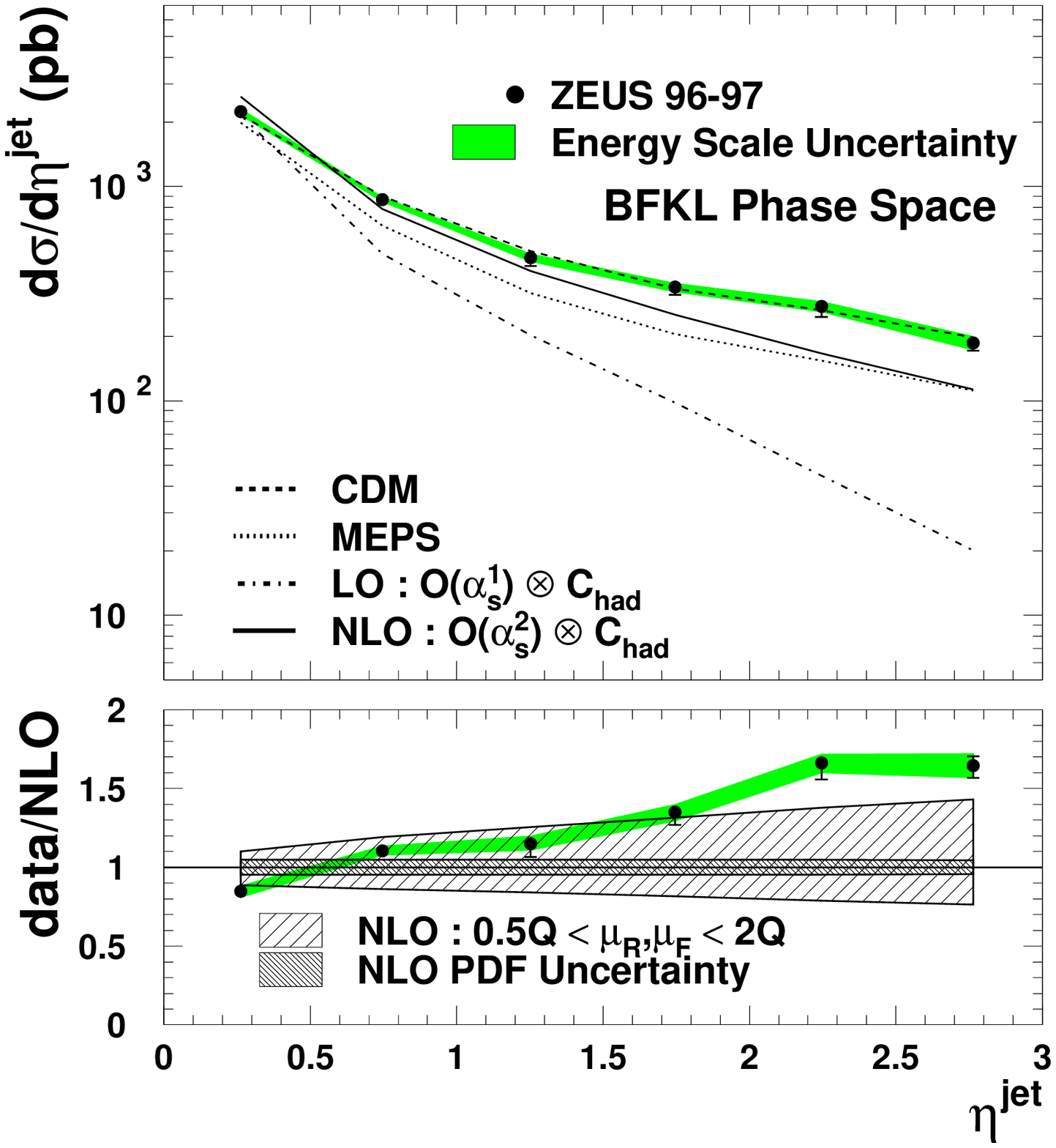,width=10cm}}
\put (8.0,9.0){\epsfig{file=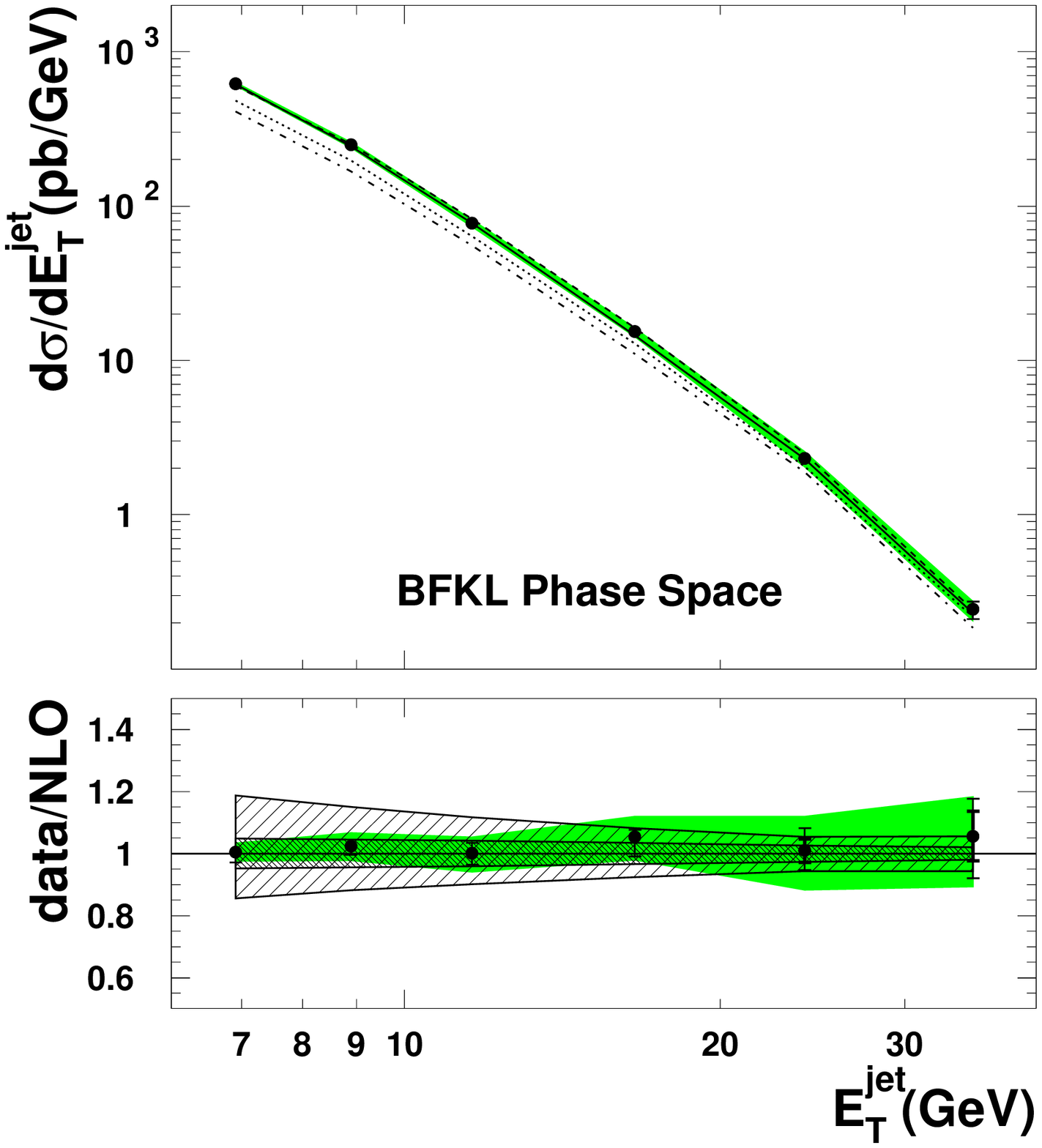,width=10cm}}
\put (-1.0,0.0){\epsfig{file=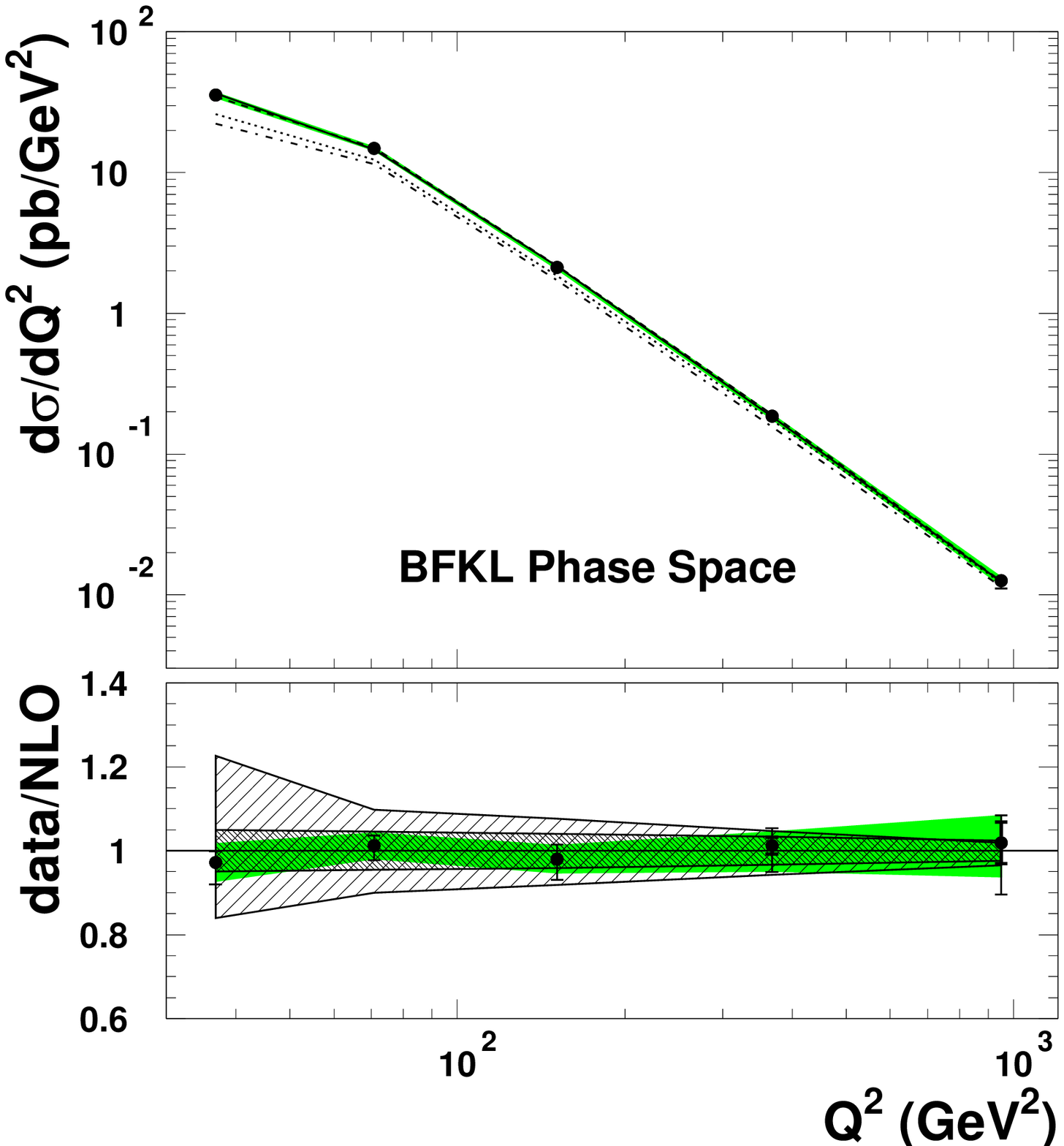,width=10cm}}
\put (8.0,0.0){\epsfig{file=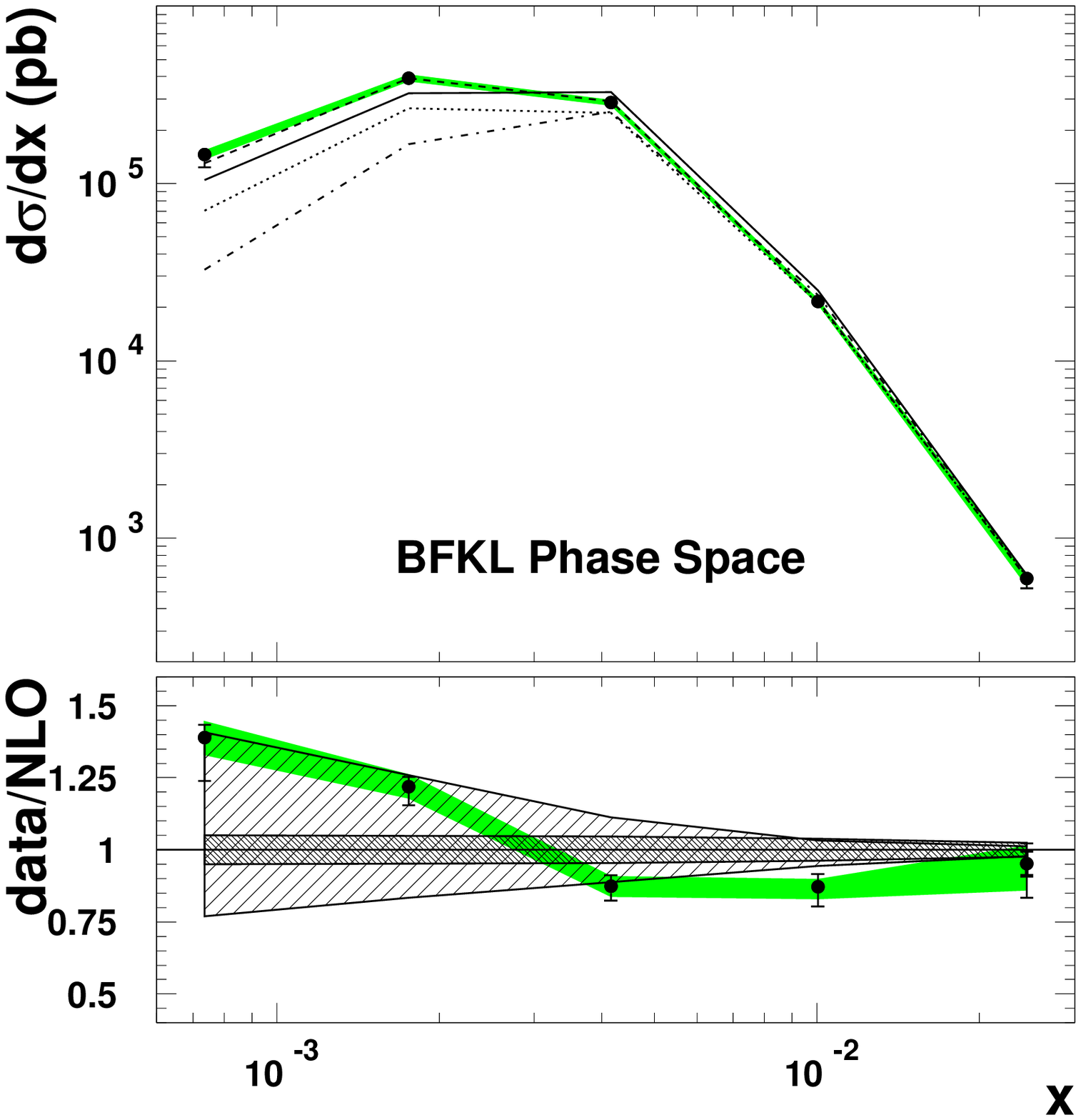,width=10cm}}
\put (5.5,13.3){\bf\small (a)}
\put (15.0,17.0){\bf\small (b)}
\put (6.0,7.5){\bf\small (c)}
\put (15.0,7.5){\bf\small (d)}
\put (7.25,19.0){\bf\huge ZEUS}
\end{picture}
\caption{
Differential cross sections (dots) in the BFKL phase space for
inclusive jet production in NC DIS with $\etjet > 6$~${\rm GeV}$,
$0 < \etajet < 3$, $\q2 > 25$~${\rm GeV}^2$, $y > 0.04$, $\cos\gh < 0$
and $0.5 < (\etjet)^2/Q^2 < 2$ as functions of (a) $\etajet$,
(b) $\etjet$, (c) $\q2$ and (d) $x$. The $\oas$ (dot-dashed lines)
and ${\cal O}(\alpha_s^2)$ (solid lines) QCD calculations are shown.
Other details are as in the caption to Fig.~2.
}
\label{fig-bfkl}
\vfill
\end{figure}

The measurements in the BFKL phase-space region are presented in
Fig.~\ref{fig-bfkl} and Tables 5 to 8. 
The shape of the cross section as a function of
$\etajet$ is steeply falling in the forward region due to the
restriction on $\gh$. The predictions of {\sc Ariadne} describe all data
distributions well. The predictions of {\sc Lepto} fail to describe
the data, especially in the $\etajet$ distribution and low-$x$ region.

Fixed-order QCD calculations, which are $\oass$ in this phase-space
region, describe the data well for $\q2$, $\etjet$ and $x$, but
underestimate the data at high values of $\etajet$.  This disagreement
is concentrated in a region where the cross section is small, and so
it is not reflected in the other distributions.  The uncertainty
of the calculations due to the absence of higher orders is larger
than before, and is a more realistic estimation than in the global
phase-space region: in the present case the calculation is 
${\cal O}(\alpha_s^2)$, which contains the first contribution from
$\hat{t}$-channel gluon-exchange diagrams (see Fig.~\ref{fig-gluon}b). 

These features were investigated by comparing the LO 
(${\cal O}(\alpha_s^1)$) and NLO (${\cal O}(\alpha_s^2)$)
calculations: a) the scale variation of the NLO calculation is reduced
with respect to that of the LO calculation (not shown) for the cross
sections, except for $d\sigma/d\etajet$ in the region $\etajet >1$ and
for $d\sigma/dx$ at low $x$, where it is larger; b) in these regions
the NLO corrections are largest and the ratio NLO/LO reaches values as
large as five for $\etajet \sim 3$. The large increase of the cross
section from LO to NLO at low $x$ and large $\etajet$ is associated
with the contribution from $\hat{t}$-channel gluon-exchange
diagrams~\cite{prl:78:428}. The sizeable scale variation at NLO arises
from the fact that such contributions come from tree-level diagrams
with three final-state partons and, as a result, the calculation
accounts in an effective way only for the lowest-order contribution
\cite{prl:78:428}. Thus, the cross-section calculations at low $x$ and
large $\etajet$ are expected to be subjected to large corrections from
higher-order terms.

\subsection{The forward BFKL phase-space region}
\begin{figure}[p]
\vfill
\setlength{\unitlength}{1.0cm}
\begin{picture} (18.0,18.0)
\put (-1.0,9.0){\epsfig{file=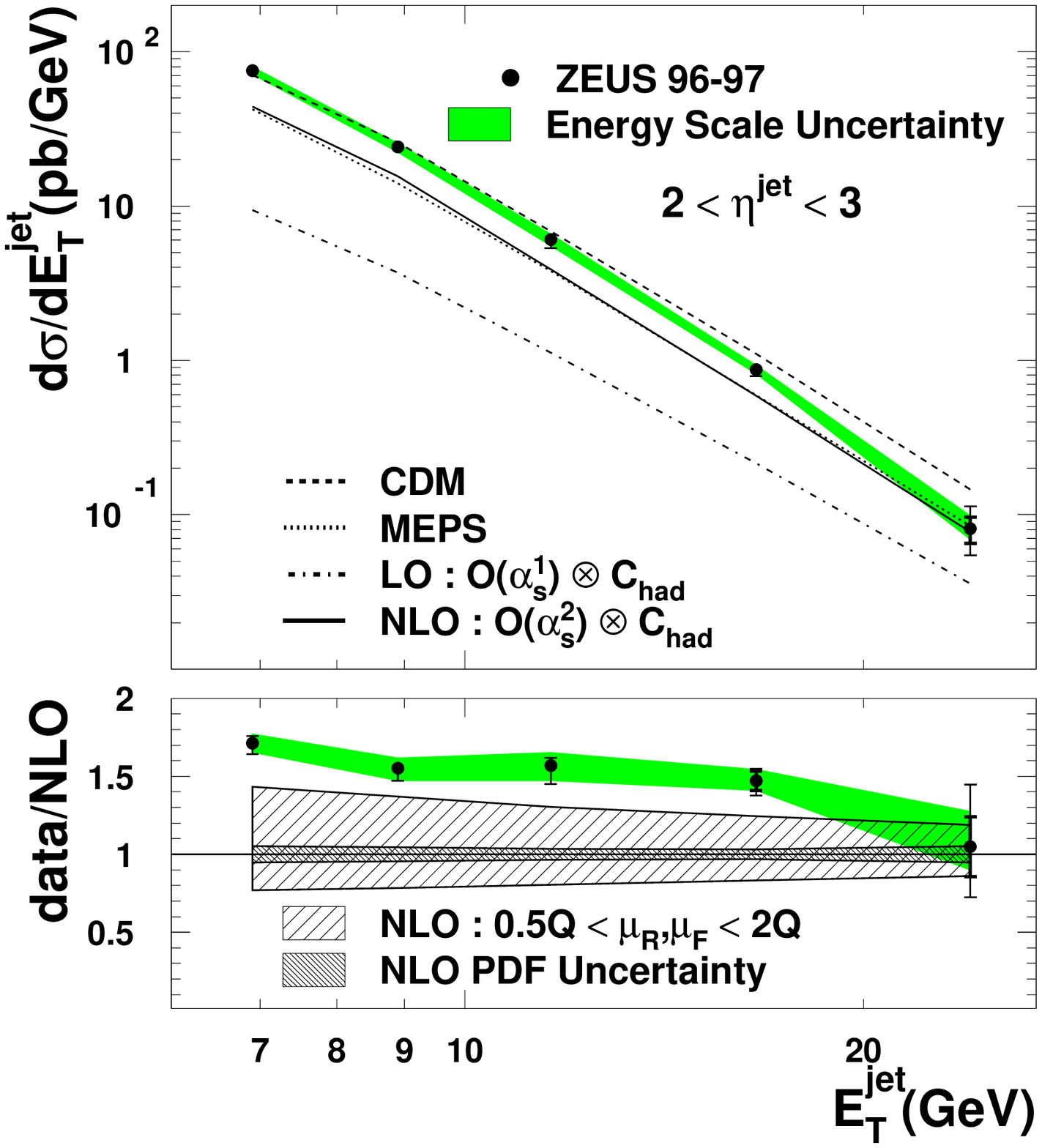,width=10cm}}
\put (8.0,9.0){\epsfig{file=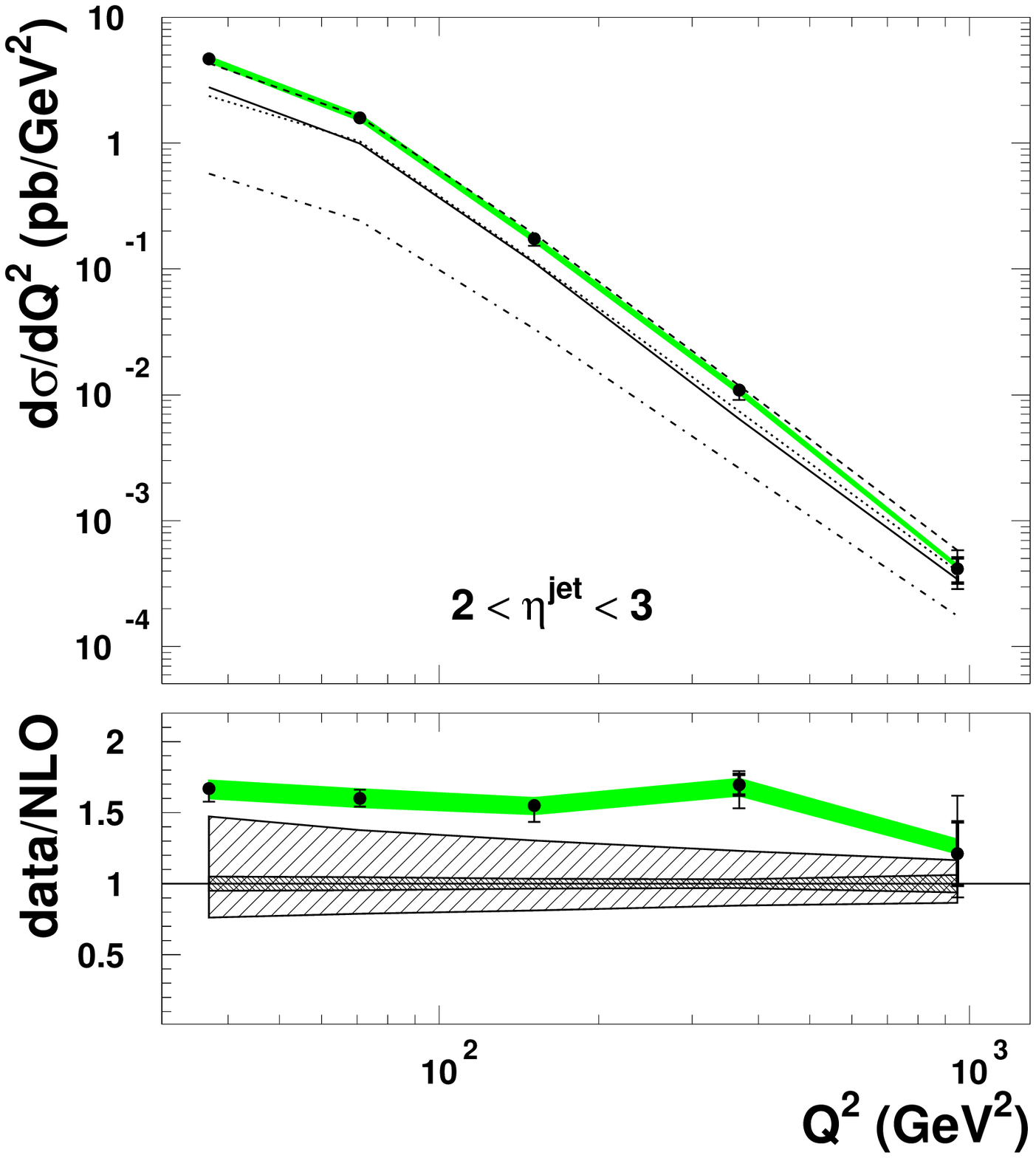,width=10cm}}
\put (2.5,-1.0){\epsfig{file=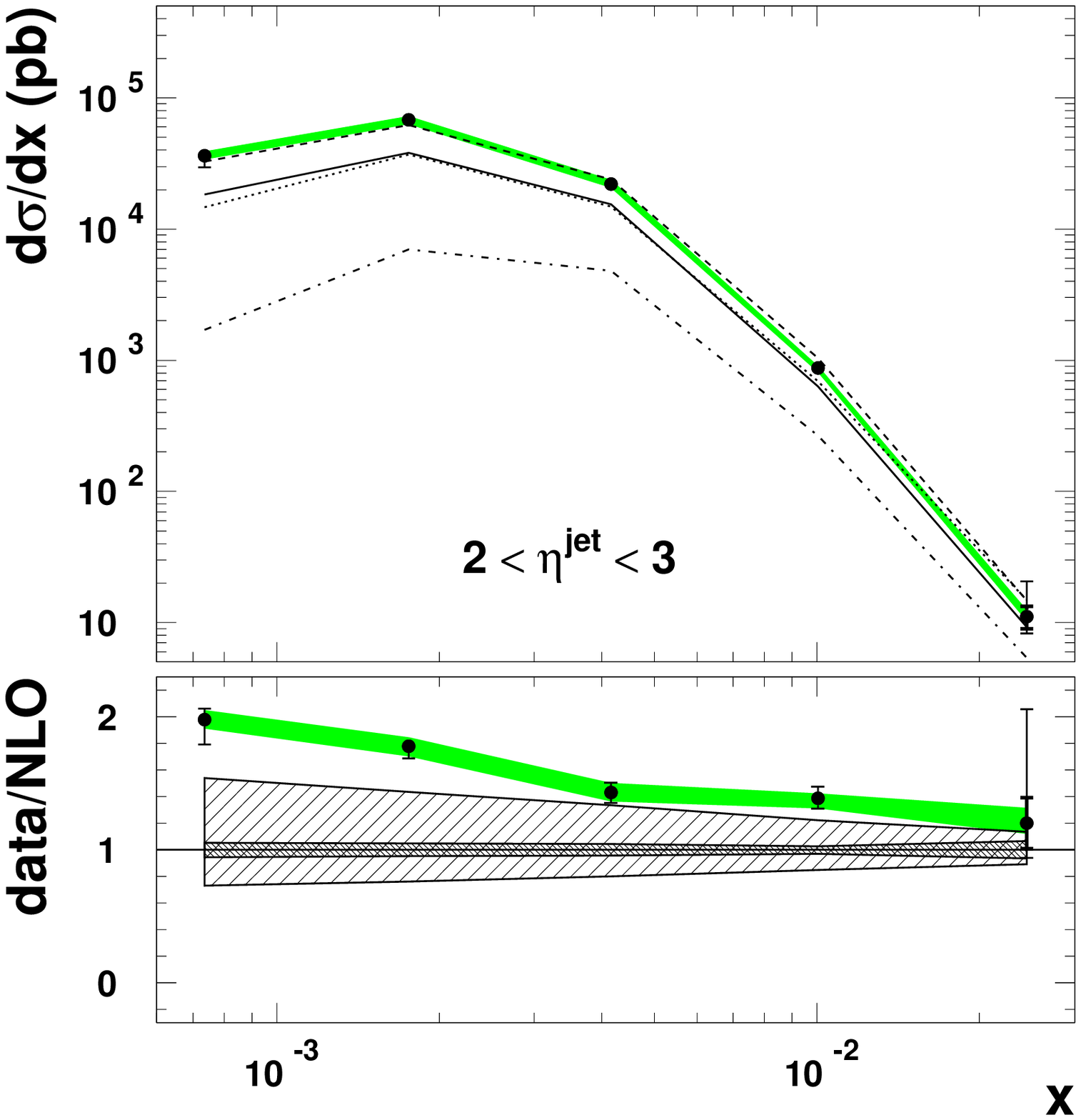,width=11cm}}
\put (6.0,15.7){\bf\small (a)}
\put (15.0,17.0){\bf\small (b)}
\put (10.3,7.2){\bf\small (c)}
\put (7.25,19.0){\bf\huge ZEUS}
\end{picture}
\vspace{1.0cm}
\caption{
Differential cross sections (dots) in the forward BFKL phase space for
inclusive jet production in NC DIS with $\etjet > 6$~${\rm GeV}$,
$2 < \etajet < 3$, $\q2 > 25$~${\rm GeV}^2$, $y > 0.04$, $\cos\gh < 0$
and $0.5 < (\etjet)^2/Q^2 < 2$ as functions of (a) $\etjet$, (b) $\q2$
and (c) $x$. The $\oas$ (dot-dashed lines) and ${\cal O}(\alpha_s^2)$
(solid lines) QCD calculations are shown. Other details are as in the
caption to Fig.~2.
}
\label{fig-forw}
\vfill
\end{figure}

The measurements in the forward BFKL phase-space region are presented in
Fig.~\ref{fig-forw} and Tables 9 to 11. 
Events exhibiting BFKL effects are expected to be dominant
in this phase-space region. The predictions of {\sc Ariadne} describe
the data well, whereas the predictions of {\sc Lepto} fail in all
distributions. Fixed-order QCD calculations are consistently lower
than the data for $\etjet$ and $Q^2$. The calculations describe the
measurement as a function of $x$ at high values, but underestimate the
data in the low-$x$ region by nearly a factor of two. The features
observed in the comparison of LO and NLO calculations in Section 8.2
are more dramatic in the present case: a) the scale variations of the
NLO calculations are larger than those of LO calculations everywhere
except at high $x$; b) the NLO corrections are large everywhere except
at high $x$ and the ratio NLO/LO reaches values as large as ten at low
$x$. The increase of the cross-section calculations from LO to NLO,
which brings the predictions closer to the data, is associated with
the contribution from $\hat{t}$-channel gluon-exchange diagrams, which
represent the lowest-order term of the expansion that leads to the
BFKL resummation. The increased scale variation at NLO, which is
larger by nearly a factor of two than that in the BFKL phase-space
region, highlights the need for improved calculations.
 
\section{Summary}
Measurements of differential cross sections in $\etjet$, $\etajet$, $Q^2$
and $x$ for inclusive jet production in neutral current deep inelastic
scattering have been presented using 38.7~${\rm pb}^{-1}$ of ZEUS data.
The low-$x$ region has been probed for events with $Q^2 > 25$~GeV$^2$
and at least one jet with $\etjet > 6$~GeV.  Three phase-space regions
have been studied: one inclusive region (global phase space), one with
an additional requirement on the hadronic angle of the event
($\cos{\gamma_h} < 0$) and a more limited window of jet pseudorapidity
$(0<\etajet<3)$, as well as the requirement 
$0.5 < (\etjet)^2/Q^2 < 2.0$ (BFKL phase space), and finally the more
restricted region with $2 < \etajet < 3$ (forward BFKL phase
space). The restrictions imposed in the BFKL phase-space regions
enhance the multijet contributions without restricting the transverse
energy of the parton(s) close to the hard scatter.
 
A large excess of the data over the fixed-order (${\cal O}(\alpha_s)$)
QCD calculation is observed in the global phase-space region at high
$\etajet$ and low $x$. This excess cannot be accomodated within the
experimental and the estimated theoretical uncertainties. However, the
size of the higher-order terms might be underestimated since the scale
variations cannot reflect the contributions from $\hat{t}$-channel
gluon-exchange diagrams, which are expected to become dominant at low
$x$.

In the BFKL phase-space region, the fixed-order (${\cal O}(\alpha_s^2)$)
QCD calculation gives, in general, a good description of the data
except for $\etajet >2$, where an excess of the data over the
prediction is observed. In this phase-space region, the NLO QCD
corrections significantly reduce the scale variation of the predicted
cross sections with respect to a LO calculation, except for
$d\sigma/d\etajet$ in the region $\etajet >1$ and for $d\sigma/dx$ at
low $x$. In these regions, the NLO corrections, which account for the
lowest-order contribution from $\hat{t}$-channel gluon-exchange
diagrams, are largest and bring the calculations close to the
data. However, the strong dependence of the calculations with the
renormalisation scale is indicative of the importance of higher-order
terms in these regions.

In the forward BFKL region, the fixed-order (${\cal O}(\alpha_s^2)$)
QCD calculation describes the shape of the measured differential cross
sections $d\sigma/d\etjet$ and $d\sigma/d\q2$, but fails to describe
that of $d\sigma/dx$. The restriction to the region $2 < \etajet < 3$
enhances the contribution from $\hat{t}$-channel gluon-exchange
diagrams, which increases the NLO prediction by up to a factor of ten
at low $x$ with respect to a LO calculation and brings it closer to
the data. The variation of the calculations with the renormalisation
scale is large, emphasizing the need for higher-order
calculations. The improved description of the data in this region
achieved by accounting for the lowest-order contribution from
$\hat{t}$-channel gluon-exchange diagrams, highlights the importance
of such terms in the parton dynamics at low $x$.

\vspace{2cm}
\noindent {\Large\bf Acknowledgments}
\vspace{1cm}

We thank the DESY Directorate for their strong support and encouragement.
We are grateful for the support of the DESY computing and network
services. The diligent efforts of the HERA machine group are
gratefully acknowledged. The design, construction and installation of
the ZEUS detector have been made possible due to the ingenuity and
efforts of many people from DESY and other institutes who are not
listed as authors.

\vfill\eject

\providecommand{\etal}{et al.\xspace}
\providecommand{\coll}{Coll.}
\catcode`\@=11
\def\@bibitem#1{%
\ifmc@bstsupport
  \mc@iftail{#1}%
    {;\newline\ignorespaces}%
    {\ifmc@first\else.\fi\orig@bibitem{#1}}
  \mc@firstfalse
\else
  \mc@iftail{#1}%
    {\ignorespaces}%
    {\orig@bibitem{#1}}%
\fi}%
\catcode`\@=12

\newpage
\clearpage
\begin{table}[p]
\begin{center}
\mbox{
\hspace{-1.8cm}
\begin{tabular}{|c|cccc||c||c|}
\hline
\raisebox{0.25cm}[1.cm]{\parbox{2cm}{\centerline{$\etajet$ bin} }}
          & \raisebox{0.25cm}[1.cm]{\parbox{2cm}{\centerline{$d\sigma/d\etajet$} \centerline{(pb)}}}
                           & $\Delta_{stat}$ &
                             $\Delta_{syst}$ &
                             $\Delta_{\text{jet}-ES}$ &
           \raisebox{0.2cm}[0.8cm]{\parbox{2.cm}{\centerline{QED} \vspace{-.2cm} \centerline{correction}}} &
           \raisebox{0.2cm}[0.8cm]{\parbox{2.8cm}{\centerline{PAR to HAD} \vspace{-0.2cm} \centerline{correction}}}\\[.05cm]
\hline
   $-1\;-\;-0.5$ & $2722$ & $\pm 13$ & $^{+25}_{-46}$  & $^{+162}_{-160}$ & $0.989$ & $0.812 \pm 0.004$   \\
   $-0.5\;-\;0$  & $3788$ & $\pm 13$ & $^{+36}_{-108}$ & $^{+157}_{-161}$ & $0.972$ & $0.842 \pm 0.001$   \\
   $0\;-\;0.5$   & $4362$ & $\pm 14$ & $^{+71}_{-82}$  & $^{+175}_{-178}$ & $0.956$ & $0.856 \pm 0.005$   \\
   $0.5\;-\;1$   & $4791$ & $\pm 15$ & $^{+38}_{-97}$  & $^{+163}_{-175}$ & $0.951$ & $0.880 \pm 0.004$   \\
   $1\;-\;1.5$   & $4217$ & $\pm 16$ & $^{+24}_{-34}$  & $^{+106}_{-109}$ & $0.947$ & $1.025 \pm 0.007$   \\
   $1.5\;-\;2$   & $2538$ & $\pm 11$ & $^{+18}_{-22}$  & $^{+71}_{-71}$   & $0.953$ & $1.12  \pm 0.03$    \\
   $2\;-\;2.5$   & $1323$ & $\pm  8$ & $^{+25}_{-132}$ & $^{+46}_{-45}$   & $0.959$ & $1.02  \pm 0.05$    \\
   $2.5\;-\;3$   &  $685$ & $\pm  5$ & $^{+36}_{-74}$  & $^{+39}_{-37}$   & $0.960$ & $0.97  \pm 0.05$    \\
\hline
\end{tabular}}
\vspace{1cm}
\caption{
Inclusive jet cross-section $d\sigma/d\etajet$ for jets of hadrons in
the global phase space. The statistical, systematic and
jet-energy-scale uncertainties are shown separately. The
multiplicative correction applied to correct for QED radiative effects
and for hadronisation effects are shown in the last two columns.
}
\label{tabGeta1}
\end{center}
\end{table}

\begin{table}[p]
\begin{center}
\mbox{
\hspace{-1.8cm}
\begin{tabular}{|c|cccc||c||c|}
\hline
\raisebox{0.25cm}[1.cm]{\parbox{2cm}{\centerline{$\etjet$ bin} \centerline{(GeV)}}}
          & \raisebox{0.25cm}[1.cm]{\parbox{2cm}{\centerline{$d\sigma/d\etjet$} \centerline{(pb/GeV)}}}
                           & $\Delta_{stat}$ &
                             $\Delta_{syst}$ &
                             $\Delta_{\text{jet}-ES}$ &
           \raisebox{0.2cm}[0.8cm]{\parbox{2.cm}{\centerline{QED} \vspace{-.2cm} \centerline{correction}}} &
           \raisebox{0.2cm}[0.8cm]{\parbox{2.8cm}{\centerline{PAR to HAD} \vspace{-0.2cm} \centerline{correction}}}\\[.05cm]
\hline
   $6\;-\;8$ & $2685$ & $\pm 6$ & $^{+20}_{-72}$ & $^{+26}_{-33}$                           & $0.965$ & $0.910  \pm 0.006$   \\
   $8\;-\;10$ & $1408$ & $\pm 4$ & $^{+8}_{-36}$ & $^{+53}_{-57}$                           & $0.954$ & $0.9163 \pm 0.0005$  \\
   $10\;-\;14$ & $599.9$ & $\pm 1.9$ & $^{+4.5}_{-9.3}$ & $^{+37.6}_{-36.3}$                & $0.957$ & $0.917  \pm 0.003$   \\
   $14\;-\;21$ & $165.55$ & $\pm 0.75$ & $^{+1.94}_{-2.40}$ & $^{+12.51}_{-11.75}$          & $0.961$ & $0.93   \pm 0.02$    \\
   $21\;-\;29$ & $40.59$ & $\pm 0.35$ & $^{+0.83}_{-0.84}$ & $^{+3.58}_{-3.82}$             & $0.956$ & $0.94   \pm 0.02$    \\
   $29\;-\;47$ & $7.90$ & $\pm 0.10$ & $^{+0.18}_{-0.21}$ & $^{+0.82}_{-0.76}$              & $0.953$ & $0.96   \pm 0.01$    \\
   $47\;-\;71$ & $0.873$ & $\pm 0.030$ & $^{+0.052}_{-0.043}$ & $^{+0.120}_{-0.095}$        & $0.966$ & $0.965  \pm 0.007$   \\
   $71\;-\;127$ & $0.0433$ & $\pm 0.0044$ & $^{+0.0080}_{-0.0047}$ & $^{+0.0068}_{-0.0100}$ & $0.996$ & $0.980  \pm 0.001$   \\
\hline
\end{tabular}}
\vspace{1cm}
\caption{
Inclusive jet cross-section $d\sigma/d\etjet$ for jets of hadrons in
the global phase space. The statistical, systematic and
jet-energy-scale uncertainties are shown separately. The
multiplicative correction applied to correct for QED radiative effects
and for hadronisation effects are shown in the last two columns.
}
\label{tabGEt1}
\end{center}
\vfill
\end{table}

\clearpage

\begin{table}[p]
\begin{center}
\mbox{
\hspace{-1.8cm}
\begin{tabular}{|c|cccc||c||c|}
\hline
\raisebox{0.25cm}[1.cm]{\parbox{2cm}{\centerline{$\q2$ bin} \centerline{(GeV$^2$)}}}
          & \raisebox{0.25cm}[1.cm]{\parbox{2cm}{\centerline{$d\sigma/dQ^2$} \centerline{(pb/GeV$^2$)}}}
                           & $\Delta_{stat}$ &
                             $\Delta_{syst}$ &
                             $\Delta_{\text{jet}-ES}$ &
           \raisebox{0.2cm}[0.8cm]{\parbox{2.cm}{\centerline{QED} \vspace{-.2cm} \centerline{correction}}} &
           \raisebox{0.2cm}[0.8cm]{\parbox{2.8cm}{\centerline{PAR to HAD} \vspace{-0.2cm} \centerline{correction}}}\\[.05cm]
\hline
   $25\;-\;50$ & $127.11$ & $\pm 0.35$ & $^{+2.06}_{-8.64}$ & $^{+8.46}_{-8.55}$                   & $0.970$ & $0.85   \pm 0.01$    \\
   $50\;-\;100$ & $73.26$ & $\pm 0.20$ & $^{+0.66}_{-1.46}$ & $^{+2.32}_{-2.47}$                   & $0.961$ & $0.937  \pm 0.008$   \\
   $100\;-\;250$ & $17.965$ & $\pm 0.055$ & $^{+0.140}_{-0.174}$ & $^{+0.139}_{-0.157}$            & $0.952$ & $1.001  \pm 0.001$   \\
   $250\;-\;630$ & $2.550$ & $\pm 0.013$ & $^{+0.019}_{-0.036}$ & $^{+0.003}_{-0.004}$             & $0.947$ & $1.0046 \pm 0.0003$  \\
   $630\;-\;1600$ & $0.3128$ & $\pm 0.0028$ & $^{+0.0039}_{-0.0049}$ & $^{+0.0002}_{-0.0003}$      & $0.934$ & $1.003  \pm 0.001$   \\
   $1600\;-\;4000$ & $0.03072$ & $\pm 0.00057$ & $^{+0.00091}_{-0.00060}$ & $^{+0}_{-0}$           & $0.930$ & $1.0004 \pm 0.0009$  \\
   $4000\;-\;10^{5}$ & $0.0001764$ & $\pm 0.0000069$ & $^{+0.0000124}_{-0.0000092}$ & $^{+0}_{-0}$ & $0.975$ & $0.998  \pm 0.002$   \\
\hline
\end{tabular}}
\vspace{1cm}
\caption{
Cross-section $d\sigma/dQ^2$ for events in the global phase space. The
statistical, systematic and jet-energy-scale uncertainties are shown
separately. The multiplicative correction applied to correct for QED
radiative effects and for hadronisation effects are shown in the last
two columns.
}
\label{tabGq21}
\end{center}
\end{table}

\begin{table}[p]
\begin{center}
\mbox{
\hspace{-1.8cm}
\begin{tabular}{|c|cccc||c||c|}
\hline
\raisebox{0.25cm}[1.cm]{\parbox{2cm}{\centerline{$x$ bin} }}
          & \raisebox{0.25cm}[1.cm]{\parbox{2cm}{\centerline{$d\sigma/dx$} \centerline{(nb)}}}
                           & $\Delta_{stat}$ &
                             $\Delta_{syst}$ &
                             $\Delta_{\text{jet}-ES}$ &
           \raisebox{0.2cm}[0.8cm]{\parbox{2.cm}{\centerline{QED} \vspace{-.2cm} \centerline{correction}}} &
           \raisebox{0.2cm}[0.8cm]{\parbox{2.8cm}{\centerline{PAR to HAD} \vspace{-0.2cm} \centerline{correction}}}\\[.05cm]
\hline
   $.0001\;-\;.001$ & $507.7$ & $\pm 3.7$ & $^{+21.4}_{-48.8}$ & $^{+28.8}_{-28.5}$       & $1.029$ & $0.94    \pm 0.03$      \\
   $.001\;-\;.0025$ & $1255.6$ & $\pm 4.6$ & $^{+19.7}_{-114.3}$ & $^{+73.8}_{-74.5}$     & $0.985$ & $0.914   \pm 0.002$     \\
   $.0025\;-\;.0063$ & $883.5$ & $\pm 2.4$ & $^{+35.6}_{-22.4}$ & $^{+40.3}_{-41.2}$      & $0.965$ & $0.903   \pm 0.004$     \\
   $.0063\;-\;.0158$ & $330.35$ & $\pm 0.93$ & $^{+2.00}_{-2.61}$ & $^{+6.11}_{-7.01}$    & $0.945$ & $0.928   \pm 0.002$     \\
   $.0158\;-\;.04$ & $60.22$ & $\pm 0.26$ & $^{+0.42}_{-3.42}$ & $^{+0.18}_{-0.19}$       & $0.942$ & $0.9899  \pm 0.0005$    \\
   $.04\;-\;.1$ & $7.607$ & $\pm 0.056$ & $^{+0.064}_{-0.408}$ & $^{+0.004}_{-0.002}$     & $0.922$ & $0.9989  \pm 0.0007$    \\
   $.1\;-\;1$ & $0.1622$ & $\pm 0.0023$ & $^{+0.0048}_{-0.0032}$ & $^{+0.0001}_{-0.0001}$ & $0.914$ & $0.99960 \pm 0.00001$   \\
\hline
\end{tabular}}
\vspace{1cm}
\caption{
Cross-section $d\sigma/dx$ for events in the global phase space. The
statistical, systematic and jet-energy-scale uncertainties are shown
separately. The multiplicative correction applied to correct for QED
radiative effects and for hadronisation effects are shown in the last
two columns.
}
\label{tabGxBj1}
\end{center}
\end{table}

\newpage
\clearpage
\begin{table}[p]
\begin{center}
\mbox{
\hspace{-1.8cm}
\begin{tabular}{|c|cccc||c||c|}
\hline
\raisebox{0.25cm}[1.cm]{\parbox{2cm}{\centerline{$\etajet$ bin} }}
          & \raisebox{0.25cm}[1.cm]{\parbox{2cm}{\centerline{$d\sigma/d\etajet$} \centerline{(pb)}}}
                           & $\Delta_{stat}$ &
                             $\Delta_{syst}$ &
                             $\Delta_{\text{jet}-ES}$ &
           \raisebox{0.2cm}[0.8cm]{\parbox{2.cm}{\centerline{QED} \vspace{-.2cm} \centerline{correction}}} &
           \raisebox{0.2cm}[0.8cm]{\parbox{2.8cm}{\centerline{PAR to HAD} \vspace{-0.2cm} \centerline{correction}}}\\[.05cm]
\hline
   $0\;-\;0.5$ & $2228$ & $\pm 11$ & $^{+41}_{-50}$ & $^{+74}_{-81}$ & $0.951$ & $1.10 \pm 0.02$   \\
   $0.5\;-\;1$ & $869$ & $\pm 7$ & $^{+22}_{-31}$ & $^{+24}_{-24}$   & $0.977$ & $1.10 \pm 0.06$   \\
   $1\;-\;1.5$ & $464$ & $\pm 5$ & $^{+13}_{-40}$ & $^{+21}_{-19}$   & $0.981$ & $1.01 \pm 0.04$   \\
   $1.5\;-\;2$ & $339$ & $\pm 4$ & $^{+10}_{-27}$ & $^{+15}_{-18}$   & $0.973$ & $0.98 \pm 0.06$   \\
   $2\;-\;2.5$ & $276$ & $\pm 4$ & $^{+9}_{-30}$ & $^{+14}_{-16}$    & $1.004$ & $0.93 \pm 0.06$   \\
   $2.5\;-\;3$ & $186$ & $\pm 3$ & $^{+11}_{-15}$ & $^{+13}_{-14}$   & $0.993$ & $0.93 \pm 0.06$   \\
\hline
\end{tabular}}
\vspace{1cm}
\caption{
Inclusive jet cross-section $d\sigma/d\etajet$ for jets of hadrons in
the BFKL phase space. The statistical, systematic and jet-energy-scale
uncertainties are shown separately. The multiplicative correction
applied to correct for QED radiative effects and for hadronisation
effects are shown in the last two columns.
}
\label{tabGeta2}
\end{center}
\end{table}

\begin{table}[p]
\begin{center}
\mbox{
\hspace{-1.8cm}
\begin{tabular}{|c|cccc||c||c|}
\hline
\raisebox{0.25cm}[1.cm]{\parbox{2cm}{\centerline{$\etjet$ bin} \centerline{(GeV)}}}
          & \raisebox{0.25cm}[1.cm]{\parbox{2cm}{\centerline{$d\sigma/d\etjet$} \centerline{(pb/GeV)}}}
                           & $\Delta_{stat}$ &
                             $\Delta_{syst}$ &
                             $\Delta_{\text{jet}-ES}$ &
           \raisebox{0.2cm}[0.8cm]{\parbox{2.cm}{\centerline{QED} \vspace{-.2cm} \centerline{correction}}} &
           \raisebox{0.2cm}[0.8cm]{\parbox{2.8cm}{\centerline{PAR to HAD} \vspace{-0.2cm} \centerline{correction}}}\\[.05cm]
\hline
   $6\;-\;8$ & $619.0$ & $\pm 2.9$ & $^{+10.3}_{-20.6}$ & $^{+17.3}_{-17.2}$         & $0.967$ & $1.06  \pm 0.05$    \\
   $8\;-\;10$ & $249.3$ & $\pm 1.8$ & $^{+4.0}_{-6.9}$ & $^{+10.8}_{-11.3}$          & $0.964$ & $1.06  \pm 0.03$    \\
   $10\;-\;14$ & $77.76$ & $\pm 0.73$ & $^{+2.64}_{-2.81}$ & $^{+4.01}_{-4.61}$      & $0.964$ & $1.08  \pm 0.03$    \\
   $14\;-\;21$ & $15.32$ & $\pm 0.25$ & $^{+0.36}_{-0.95}$ & $^{+1.00}_{-1.11}$      & $0.976$ & $1.04  \pm 0.04$    \\
   $21\;-\;29$ & $2.325$ & $\pm 0.089$ & $^{+0.170}_{-0.147}$ & $^{+0.254}_{-0.292}$ & $0.951$ & $1.052 \pm 0.009$   \\
   $29\;-\;47$ & $0.245$ & $\pm 0.019$ & $^{+0.030}_{-0.033}$ & $^{+0.031}_{-0.040}$ & $0.957$ & $0.951 \pm 0.004$   \\
\hline
\end{tabular}}
\vspace{1cm}
\caption{
Inclusive jet cross-section $d\sigma/d\etjet$ for jets of hadrons in
the BFKL phase space. The statistical, systematic and jet-energy-scale
uncertainties are shown separately. The multiplicative correction
applied to correct for QED radiative effects and for hadronisation
effects are shown in the last two columns.
}
\label{tabGEt2}
\end{center}
\vfill
\end{table}

\clearpage

\begin{table}[p]
\begin{center}
\mbox{
\hspace{-1.8cm}
\begin{tabular}{|c|cccc||c||c|}
\hline
\raisebox{0.25cm}[1.cm]{\parbox{2cm}{\centerline{$\q2$ bin} \centerline{(GeV$^2$)}}}
          & \raisebox{0.25cm}[1.cm]{\parbox{2cm}{\centerline{$d\sigma/dQ^2$} \centerline{(pb/GeV$^2$)}}}
                           & $\Delta_{stat}$ &
                             $\Delta_{syst}$ &
                             $\Delta_{\text{jet}-ES}$ &
           \raisebox{0.2cm}[0.8cm]{\parbox{2.cm}{\centerline{QED} \vspace{-.2cm} \centerline{correction}}} &
           \raisebox{0.2cm}[0.8cm]{\parbox{2.8cm}{\centerline{PAR to HAD} \vspace{-0.2cm} \centerline{correction}}}\\[.05cm]
\hline
   $25\;-\;50$ & $35.62$ & $\pm 0.20$ & $^{+0.98}_{-1.84}$ & $^{+1.60}_{-1.58}$                     & $0.967$ & $1.01  \pm 0.03$   \\
   $50\;-\;100$ & $14.858$ & $\pm 0.090$ & $^{+0.357}_{-0.528}$ & $^{+0.431}_{-0.463}$              & $0.968$ & $1.12  \pm 0.04$   \\
   $100\;-\;250$ & $2.121$ & $\pm 0.020$ & $^{+0.074}_{-0.106}$ & $^{+0.068}_{-0.070}$              & $0.954$ & $1.15  \pm 0.03$   \\
   $250\;-\;630$ & $0.1868$ & $\pm 0.0038$ & $^{+0.0077}_{-0.0118}$ & $^{+0.0058}_{-0.0112}$        & $0.985$ & $1.16  \pm 0.02$   \\
   $630\;-\;1600$ & $0.01262$ & $\pm 0.00062$ & $^{+0.00082}_{-0.00155}$ & $^{+0.00080}_{-0.00102}$ & $0.957$ & $1.144 \pm 0.003$   \\
\hline
\end{tabular}}
\vspace{1cm}
\caption{
Cross-section $d\sigma/dQ^2$ for events in the BFKL phase space. The
statistical, systematic and jet-energy-scale uncertainties are shown
separately. The multiplicative correction applied to correct for QED
radiative effects and for hadronisation effects are shown in the last
two columns.
}
\label{tabGq22}
\end{center}
\end{table}

\begin{table}[p]
\begin{center}
\mbox{
\hspace{-1.8cm}
\begin{tabular}{|c|cccc||c||c|}
\hline
\raisebox{0.25cm}[1.cm]{\parbox{2cm}{\centerline{$x$ bin} }}
          & \raisebox{0.25cm}[1.cm]{\parbox{2cm}{\centerline{$d\sigma/dx$} \centerline{(nb)}}}
                           & $\Delta_{stat}$ &
                             $\Delta_{syst}$ &
                             $\Delta_{\text{jet}-ES}$ &
           \raisebox{0.2cm}[0.8cm]{\parbox{2.cm}{\centerline{QED} \vspace{-.2cm} \centerline{correction}}} &
           \raisebox{0.2cm}[0.8cm]{\parbox{2.8cm}{\centerline{PAR to HAD} \vspace{-0.2cm} \centerline{correction}}}\\[.05cm]
\hline
   $.0001\;-\;.001$ & $145.6$ & $\pm 1.9$ & $^{+6.2}_{-22.1}$ & $^{+7.9}_{-8.5}$         & $1.023$ & $0.90   \pm 0.05$     \\
   $.001\;-\;.0025$ & $393.6$ & $\pm 2.6$ & $^{+13.2}_{-25.0}$ & $^{+15.8}_{-14.8}$      & $0.992$ & $1.01   \pm 0.05$     \\
   $.0025\;-\;.0063$ & $286.4$ & $\pm 1.5$ & $^{+10.9}_{-14.0}$ & $^{+9.6}_{-9.8}$       & $0.951$ & $1.11   \pm 0.02$     \\
   $.0063\;-\;.0158$ & $21.69$ & $\pm 0.24$ & $^{+0.98}_{-1.46}$ & $^{+0.55}_{-0.88}$    & $0.942$ & $1.2282 \pm 0.0001$   \\
   $.0158\;-\;.04$ & $0.591$ & $\pm 0.025$ & $^{+0.041}_{-0.070}$ & $^{+0.035}_{-0.054}$ & $0.944$ & $1.18   \pm 0.01$     \\
\hline
\end{tabular}}
\vspace{1cm}
\caption{
Cross-section $d\sigma/dx$ for events in the BFKL phase space. The
statistical, systematic and jet-energy-scale uncertainties are shown
separately. The multiplicative correction applied to correct for QED
radiative effects and for hadronisation effects are shown in the last
two columns.
}
\label{tabGxBj2}
\end{center}
\end{table}

\newpage
\clearpage
\begin{table}[p]
\begin{center}
\mbox{
\hspace{-1.8cm}
\begin{tabular}{|c|cccc||c||c|}
\hline
\raisebox{0.25cm}[1.cm]{\parbox{2cm}{\centerline{$\etjet$ bin} \centerline{(GeV)}}}
          & \raisebox{0.25cm}[1.cm]{\parbox{2cm}{\centerline{$d\sigma/d\etjet$} \centerline{(pb/GeV)}}}
                           & $\Delta_{stat}$ &
                             $\Delta_{syst}$ &
                             $\Delta_{\text{jet}-ES}$ &
           \raisebox{0.2cm}[0.8cm]{\parbox{2.cm}{\centerline{QED} \vspace{-.2cm} \centerline{correction}}} &
           \raisebox{0.2cm}[0.8cm]{\parbox{2.8cm}{\centerline{PAR to HAD} \vspace{-0.2cm} \centerline{correction}}}\\[.05cm]
\hline
   $6\;-\;8$ & $75.5$ & $\pm 1.0$ & $^{+3.4}_{-5.3}$ & $^{+4.0}_{-4.3}$              & $0.997$ & $0.91 \pm 0.06$   \\
   $8\;-\;10$ & $24.08$ & $\pm 0.57$ & $^{+0.81}_{-1.94}$ & $^{+1.58}_{-1.79}$       & $0.991$ & $0.95 \pm 0.06$   \\
   $10\;-\;14$ & $6.08$ & $\pm 0.20$ & $^{+0.29}_{-0.73}$ & $^{+0.49}_{-0.58}$       & $0.999$ & $0.99 \pm 0.03$   \\
   $14\;-\;21$ & $0.871$ & $\pm 0.055$ & $^{+0.067}_{-0.083}$ & $^{+0.062}_{-0.052}$ & $0.998$ & $1.03 \pm 0.07$   \\
   $21\;-\;29$ & $0.081$ & $\pm 0.016$ & $^{+0.032}_{-0.026}$ & $^{+0.018}_{-0.012}$ & $0.922$ & $1.01 \pm 0.03$   \\
\hline
\end{tabular}}
\vspace{1cm}
\caption{
Inclusive jet cross-section $d\sigma/d\etjet$ for jets of hadrons in
the forward-BFKL phase space. The statistical, systematic and
jet-energy-scale uncertainties are shown separately. The
multiplicative correction applied to correct for QED radiative effects
and for hadronisation effects are shown in the last two columns.
}
\label{tabGEt3}
\end{center}
\vfill
\end{table}

\clearpage

\begin{table}[p]
\begin{center}
\mbox{
\hspace{-1.8cm}
\begin{tabular}{|c|cccc||c||c|}
\hline
\raisebox{0.25cm}[1.cm]{\parbox{2cm}{\centerline{$\q2$ bin} \centerline{(GeV$^2$)}}}
          & \raisebox{0.25cm}[1.cm]{\parbox{2cm}{\centerline{$d\sigma/dQ^2$} \centerline{(pb/GeV$^2$)}}}
                           & $\Delta_{stat}$ &
                             $\Delta_{syst}$ &
                             $\Delta_{\text{jet}-ES}$ &
           \raisebox{0.2cm}[0.8cm]{\parbox{2.cm}{\centerline{QED} \vspace{-.2cm} \centerline{correction}}} &
           \raisebox{0.2cm}[0.8cm]{\parbox{2.8cm}{\centerline{PAR to HAD} \vspace{-0.2cm} \centerline{correction}}}\\[.05cm]
\hline
   $25\;-\;50$ & $4.650$ & $\pm 0.075$ & $^{+0.190}_{-0.420}$ & $^{+0.273}_{-0.320}$               & $0.992$ & $0.93 \pm 0.06$   \\
   $50\;-\;100$ & $1.587$ & $\pm 0.029$ & $^{+0.098}_{-0.092}$ & $^{+0.106}_{-0.094}$              & $0.999$ & $0.93 \pm 0.06$   \\
   $100\;-\;250$ & $0.1735$ & $\pm 0.0053$ & $^{+0.0070}_{-0.0202}$ & $^{+0.0094}_{-0.0108}$       & $0.984$ & $0.96 \pm 0.02$   \\
   $250\;-\;630$ & $0.01093$ & $\pm 0.00080$ & $^{+0.00103}_{-0.00180}$ & $^{+0.00047}_{-0.00083}$ & $1.010$ & $1.03 \pm 0.01$   \\
   $630\;-\;1600$ & $0.000414$ & $\pm 0.000093$ & $^{+0.000170}_{-0.000126}$ & $^{+0}_{-0.000041}$ & $0.944$ & $1.04 \pm 0.02$   \\
\hline
\end{tabular}}
\vspace{1cm}
\caption{
Cross-section $d\sigma/dQ^2$ for events in the forward-BFKL phase
space. The statistical, systematic and jet-energy-scale uncertainties
are shown separately. The multiplicative correction applied to correct
for QED radiative effects and for hadronisation effects are shown in
the last two columns.
}
\label{tabGq23}
\end{center}
\end{table}

\begin{table}[p]
\begin{center}
\mbox{
\hspace{-1.8cm}
\begin{tabular}{|c|cccc||c||c|}
\hline
\raisebox{0.25cm}[1.cm]{\parbox{2cm}{\centerline{$x$ bin} }}
          & \raisebox{0.25cm}[1.cm]{\parbox{2cm}{\centerline{$d\sigma/dx$} \centerline{(nb)}}}
                           & $\Delta_{stat}$ &
                             $\Delta_{syst}$ &
                             $\Delta_{\text{jet}-ES}$ &
           \raisebox{0.2cm}[0.8cm]{\parbox{2.cm}{\centerline{QED} \vspace{-.2cm} \centerline{correction}}} &
           \raisebox{0.2cm}[0.8cm]{\parbox{2.8cm}{\centerline{PAR to HAD} \vspace{-0.2cm} \centerline{correction}}}\\[.05cm]
\hline
   $.0001\;-\;.001$ & $36.4$ & $\pm 1.1$ & $^{+3.0}_{-6.7}$ & $^{+2.4}_{-2.2}$             & $1.066$ & $0.88 \pm 0.06$   \\
   $.001\;-\;.0025$ & $68.0$ & $\pm 1.1$ & $^{+2.5}_{-6.2}$ & $^{+4.1}_{-4.7}$             & $1.024$ & $0.91 \pm 0.07$   \\
   $.0025\;-\;.0063$ & $22.12$ & $\pm 0.41$ & $^{+1.67}_{-1.70}$ & $^{+1.46}_{-1.34}$      & $0.962$ & $0.96 \pm 0.04$   \\
   $.0063\;-\;.0158$ & $0.872$ & $\pm 0.039$ & $^{+0.077}_{-0.066}$ & $^{+0.026}_{-0.059}$ & $0.957$ & $1.00 \pm 0.03$   \\
   $.0158\;-\;.04$ & $0.01113$ & $\pm 0.0021$ & $^{+0.0095}_{-0.0029}$ & $^{+0}_{-0.0012}$ & $0.945$ & $0.9  \pm 0.1$   \\
\hline
\end{tabular}}
\vspace{1cm}
\caption{
Cross-section $d\sigma/dx$ for events in the forward-BFKL phase
space. The statistical, systematic and jet-energy-scale uncertainties
are shown separately. The multiplicative correction applied to correct
for QED radiative effects and for hadronisation effects are shown in
the last two columns.
}
\label{tabGxBj3}
\end{center}
\end{table}

\newpage
\clearpage
\pagebreak

\pagebreak

\pagebreak


\begin{mcbibliography}{10}

\bibitem{sovjnp:15:438}
V.N.~Gribov and L.N.~Lipatov,
\newblock Sov.\ J.\ Nucl.\ Phys.{} 15~(1972)~438\relax
\relax
\bibitem{sovjnp:20:94}
L.N.~Lipatov,
\newblock Sov.\ J.\ Nucl.\ Phys.{} 20~(1975)~94\relax
\relax
\bibitem{jetp:46:641}
Yu.L.~Dokshitzer,
\newblock Sov.\ Phys.\ JETP{} 46~(1977)~641\relax
\relax
\bibitem{np:b126:298}
G.~Altarelli and G.~Parisi,
\newblock Nucl.\ Phys.{} B~126~(1977)~298\relax
\relax
\bibitem{jetp:45:199}
E.A.~Kuraev, L.N.~Lipatov and V.S.~Fadin,
\newblock Sov.\ Phys.\ JETP{} 45~(1977)~199\relax
\relax
\bibitem{sovjnp:28:822}
Ya.Ya.~Balitski\u i and L.N.~Lipatov,
\newblock Sov.\ J.\ Nucl.\ Phys.{} 28~(1978)~822\relax
\relax
\bibitem{desy-03-214}
ZEUS \coll, S.~Chekanov \etal,
\newblock Phys.\ Rev.{} D~70~(2004)~052001\relax
\relax
\bibitem{pl:b547:164}
ZEUS \coll, S.~Chekanov \etal,
\newblock Phys.\ Lett.{} B~547~(2002)~164\relax
\relax
\bibitem{pr:d67:012007}
ZEUS \coll, S.~Chekanov \etal,
\newblock Phys.\ Rev.{} D~67~(2003)~012007\relax
\relax
\bibitem{pl:b507:70}
ZEUS \coll, J.~Breitweg \etal,
\newblock Phys.\ Lett.{} B~507~(2001)~70\relax
\relax
\bibitem{epj:c30:1}
H1 \coll, C.~Adloff \etal,
\newblock Eur.\ Phys.\ J.{} C~30~(2003)~1\relax
\relax
\bibitem{pl:b542:193}
H1 \coll, C.~Adloff \etal,
\newblock Phys.\ Lett.{} B~542~(2002)~193\relax
\relax
\bibitem{pl:b515:17}
H1 \coll, C.~Adloff \etal,
\newblock Phys.\ Lett.{} B~515~(2001)~17\relax
\relax
\bibitem{np:c18:125}
A.H.~Mueller,
\newblock Nucl.~Phys.~Proc.~Suppl.{} C~18~(1991)~125\relax
\relax
\bibitem{jp:g17:1443}
A.H.~Mueller,
\newblock J.\ Phys.{} G~17~(1991)~1443\relax
\relax
\bibitem{zfp:c54:645}
J.~Bartels, A.~De Roeck and M.~Loewe,
\newblock Z.\ Phys.{} C~54~(1992)~645\relax
\relax
\bibitem{pl:b287:254}
J.~Kwiecinski, A.D.~Martin and P.J.~Sutton,
\newblock Phys.\ Lett.{} B~287~(1992)~254\relax
\relax
\bibitem{pl:b278:363}
W.K.~Tang,
\newblock Phys.\ Lett.{} B~278~(1992)~363\relax
\relax
\bibitem{epj:c6:239}
ZEUS \coll, J.~Breitweg \etal,
\newblock Eur.\ Phys.\ J.{} C~6~(1999)~239\relax
\relax
\bibitem{pl:b474:223}
ZEUS \coll, J.~Breitweg \etal,
\newblock Phys.\ Lett.{} B~474~(2000)~223\relax
\relax
\bibitem{np:b538:3}
H1 \coll, C.~Adloff \etal,
\newblock Nucl.\ Phys.{} B~538~(1999)~3\relax
\relax
\bibitem{cpc:86:147}
H.~Jung,
\newblock Comp.\ Phys.\ Comm.{} 86~(1995)~147\relax
\relax
\bibitem{pl:b165:147}
Y.~Azimov \etal,
\newblock Phys.\ Lett.{} B~165~(1985)~147\relax
\relax
\bibitem{pl:b175:453}
G.~Gustafson,
\newblock Phys.\ Lett.{} B~175~(1986)~453\relax
\relax
\bibitem{np:b306:746}
G.~Gustafson and U.~Pettersson,
\newblock Nucl.\ Phys.{} B~306~(1988)~746\relax
\relax
\bibitem{zfp:c43:625}
B.~Andersson \etal,
\newblock Z.\ Phys.{} C~43~(1989)~625\relax
\relax
\bibitem{np:b406:187}
S.Catani \etal,
\newblock Nucl.\ Phys.{} B406~(1993)~187\relax
\relax
\bibitem{pr:d48:3160}
S.D.~Ellis and D.E.~Soper,
\newblock Phys.\ Rev.{} D~48~(1993)~3160\relax
\relax
\bibitem{feynman:1972:photon}
R.P.~Feynman,
\newblock {\em Photon-Hadron Interactions}.
\newblock Benjamin, New York, (1972)\relax
\relax
\bibitem{jp:g25:1473}
B.~Poetter and M.H.~Seymour,
\newblock J.\ Phys.{} G~25~(1999)~1473\relax
\relax
\bibitem{pl:b293:465}
ZEUS \coll, M.~Derrick \etal,
\newblock Phys.\ Lett.{} B~293~(1992)~465\relax
\relax
\bibitem{zeus:1993:bluebook}
\colab{ZEUS}, U.~Holm~(ed.),
\newblock {\em The {ZEUS} Detector}.
\newblock Status Report (unpublished), DESY (1993),
\newblock available on
  \texttt{http://www-zeus.desy.de/bluebook/bluebook.html}\relax
\relax
\bibitem{nim:a279:290}
N.~Harnew \etal,
\newblock Nucl.\ Inst.\ Meth.{} A~279~(1989)~290\relax
\relax
\bibitem{npps:b32:181}
B.~Foster \etal,
\newblock Nucl.\ Phys.\ Proc.\ Suppl.{} B~32~(1993)~181\relax
\relax
\bibitem{nim:a338:254}
B.~Foster \etal,
\newblock Nucl.\ Inst.\ Meth.{} A~338~(1994)~254\relax
\relax
\bibitem{nim:a309:77}
M.~Derrick \etal,
\newblock Nucl.\ Inst.\ Meth.{} A~309~(1991)~77\relax
\relax
\bibitem{nim:a309:101}
A.~Andresen \etal,
\newblock Nucl.\ Inst.\ Meth.{} A~309~(1991)~101\relax
\relax
\bibitem{nim:a321:356}
A.~Caldwell \etal,
\newblock Nucl.\ Inst.\ Meth.{} A~321~(1992)~356\relax
\relax
\bibitem{nim:a336:23}
A.~Bernstein \etal,
\newblock Nucl.\ Inst.\ Meth.{} A~336~(1993)~23\relax
\relax
\bibitem{desy-92-066}
J.~Andruszk\'ow \etal,
\newblock Preprint \mbox{DESY-92-066}, DESY, 1992\relax
\relax
\bibitem{zfp:c63:391}
ZEUS \coll, M.~Derrick \etal,
\newblock Z.\ Phys.{} C~63~(1994)~391\relax
\relax
\bibitem{acpp:b32:2025}
J.~Andruszk\'ow \etal,
\newblock Acta Phys.\ Pol.{} B~32~(2001)~2025\relax
\relax
\bibitem{hep-ex-0208037}
ZEUS \coll, S.~Chekanov \etal,
\newblock Preprint \mbox{hep-ex/0208037}, 2002\relax
\relax
\bibitem{proc:epfacility:1979:391}
F.~Jacquet and A.~Blondel,
\newblock {\em Proc. of the Study for an $ep$ Facility for {Europe}},
  U.~Amaldi~(ed.), p.~391.
\newblock Hamburg, Germany (1979).
\newblock Also in preprint \mbox{DESY 79/48}\relax
\relax
\bibitem{proc:hera:1991:23}
S.~Bentvelsen, J.~Engelen and P.~Kooijman,
\newblock {\em Proc.\ Workshop on Physics at {HERA}}, W.~Buchm\"uller and
  G.~Ingelman~(eds.), Vol.~1, p.~23.
\newblock Hamburg, Germany, DESY (1992)\relax
\relax
\bibitem{proc:snowmass:1990:134}
J.E.~Huth \etal,
\newblock {\em Research Directions for the Decade. Proceedings of Summer Study
  on High Energy Physics, 1990}, E.L.~Berger~(ed.), p.~134.
\newblock World Scientific (1992).
\newblock Also in preprint \mbox{FERMILAB-CONF-90-249-E}\relax
\relax
\bibitem{tech:cern-dd-ee-84-1}
R.~Brun et al.,
\newblock {\em {\sc geant3}},
\newblock Technical Report CERN-DD/EE/84-1, CERN, 1987\relax
\relax
\bibitem{cpc:101:108}
G.~Ingelman, A.~Edin and J.~Rathsman,
\newblock Comp.\ Phys.\ Comm.{} 101~(1997)~108\relax
\relax
\bibitem{cpc:69:155}
A.~Kwiatkowski, H.~Spiesberger and H.-J.~M\"ohring,
\newblock Comp.\ Phys.\ Comm.{} 69~(1992)~155\relax
\relax
\bibitem{cpc:81:381}
K.~Charchula, G.A.~Schuler and H.~Spiesberger,
\newblock Comp.\ Phys.\ Comm.{} 81~(1994)~381\relax
\relax
\bibitem{spi:www:djangoh11}
H.~Spiesberger,
\newblock {\em {\sc heracles} and {\sc djangoh}: Event Generation for $ep$
  Interactions at {HERA} Including Radiative Processes}, 1998,
\newblock available on \texttt{http://www.desy.de/\til
  hspiesb/djangoh.html}\relax
\relax
\bibitem{cpc:71:15}
L.~L\"onnblad,
\newblock Comp.\ Phys.\ Comm.{} 71~(1992)~15\relax
\relax
\bibitem{zp:c65:285}
L.~L\"onnblad,
\newblock Z.\ Phys.{} C~65~(1995)~285\relax
\relax
\bibitem{epj:c11:251}
ZEUS \coll, J.~Breitweg \etal,
\newblock Eur.\ Phys.\ J.{} C~11~(1999)~251\relax
\relax
\bibitem{prep:97:31}
B.~Andersson \etal,
\newblock Phys.\ Rep.{} 97~(1983)~31\relax
\relax
\bibitem{cpc:82:74}
T.~Sj\"ostrand,
\newblock Comp.\ Phys.\ Comm.{} 82~(1994)~74\relax
\relax
\bibitem{pr:d55:1280}
H.L.~Lai \etal,
\newblock Phys.\ Rev.{} D~55~(1997)~1280\relax
\relax
\bibitem{thesis:lammers:2004}
S.~Lammers.
\newblock Ph.D.\ Thesis, University of Wisconsin-Madison (2004)
  (unpublished)\relax
\relax
\bibitem{epj:c23:13}
ZEUS \coll, S.~Chekanov et al.,
\newblock Eur.\ Phys.\ J.{} C~23~(2002)~13\relax
\relax
\bibitem{np:b485:291}
S.~Catani and M.H.~Seymour,
\newblock Nucl.\ Phys.{} B~485,~291 (1997). Erratum in Nucl.~Phys.{\bf~B~510},
  503 (1998)\relax
\relax
\bibitem{hepph0201195}
J.~Pumplin \etal,
\newblock JHEP{} 0207~(2002)~012\relax
\relax
\bibitem{Graudenz:1997sp}
D.~Graudenz,
\newblock Preprint \mbox{hep-ph/9708362}, 1997\relax
\relax
\bibitem{Graudenz:1997gv}
D.~Graudenz,
\newblock Preprint \mbox{hep-ph/9710244}, 1997\relax
\relax
\bibitem{DelDuca:hepph9707348}
V.~Del Duca,
\newblock Preprint \mbox{hep-ph/9707348}, 1997\relax
\relax
\bibitem{prl:78:428}
E.~Mirkes and D.~Zeppenfeld,
\newblock Phys. Rev. Lett.{} 78~(1997)~428\relax
\relax
\end{mcbibliography}
\end{document}